\documentclass[a4paper,11pt]{article}
\pdfoutput=1
\usepackage{jheppub}
\usepackage[utf8]{inputenc}
\usepackage[english]{babel}
\usepackage[dvipsnames,svgnames,table]{xcolor}
\usepackage{graphicx}
\usepackage{epstopdf}
\usepackage{amsmath}
\usepackage{amssymb}
\usepackage{mathtools}
\usepackage{slashed}
\usepackage{braket}
\usepackage{booktabs}
\usepackage[toc,page]{appendix}
\usepackage{url}
\usepackage{nameref}

\newcommand{\eps}{\varepsilon}
\newcommand{\IR}{\scriptscriptstyle \mathrm{IR}}
\newcommand{\UV}{\scriptscriptstyle \mathrm{UV}}
\def\nbslash{\rlap{\hspace{0.02cm}/}{\bar n}}
\def\nslash{\rlap{\hspace{0.02cm}/}{n}}

\usepackage[sort&compress,numbers]{natbib}
\usepackage{doi}
\usepackage{hyperref}
\makeatletter
\AtBeginDocument{\let\NAT@space\relax}
\makeatother

\usepackage[capitalise]{cleveref}

\graphicspath{ {./Figures/} }

\title{Factorisation theorem and pole cancellation in the resolved \texorpdfstring{$Q_1^q$--$Q_{7\gamma}$}{Q1-Q7gamma} contribution to \texorpdfstring{$\bar{B}\to X_s\gamma$}{B -> Xs gamma} at NLO}

\author{Riccardo Bartocci$^{\,a,b}$ and Tobias Hurth$^{\,c}$} 
 
\emailAdd{riccardo.bartocci@kit.edu}
\emailAdd{hurth@uni-mainz.de}

\affiliation{${}^a$Institute for Theoretical Physics, Karlsruhe Institute of Technology, Wolfgang-Gaede-Str. 1, 76131 Karlsruhe,
Germany}
\affiliation{${}^b$Institute for Astroparticle Physics, Karlsruhe Institute of Technology, 76344 Eggenstein-Leopoldshafen,
Germany}
\affiliation{${}^c$PRISMA+ Cluster of Excellence \& Institute of  Physics (THEP) \& Mainz Institute for Theoretical Physics, Johannes Gutenberg University, D-55099 Mainz, Germany}

\abstract{We study the subleading resolved $Q_1^q$–$Q_{7\gamma}$ contribution to the inclusive decay $\bar{B}\to X_s\gamma$ at next-to-leading order (NLO) in $\alpha_s$.
We present a formulation of the factorisation theorem which is valid beyond the leading order (LO). On this basis, we explicitly demonstrate the cancellation of all $1/\eps$ singularities between the hard, hard-collinear, anti-hard-collinear and soft contributions.
For this purpose, a central result of this work is the computation of the $u$-integrated massless two-loop anti-hard-collinear jet function, which determines the structure of the $1/\eps$ singularities.
In addition, we compute the hard function at the required order, combine our results with the hard-collinear jet function known from the literature, and employ the shape function whose renormalisation was determined in our previous work. There, we showed, for inclusive as well as for exclusive modes, that certain contributions to these multi-light-cone soft functions are physically irrelevant and can be consistently omitted; the explicit cancellation of the corresponding divergences provides a non-trivial confirmation of this statement. 
Overall, our findings constitute a non-trivial consistency check of the factorisation framework for this subleading resolved contribution at NLO and provide the basis for a full phenomenological analysis.
}

\preprint{
\begin{minipage}{4cm}
\small
  \flushright
MITP-26-054 \\
P3H-26-085 \\
KA-TP-28-2026
\end{minipage}}

\begin{document}

\maketitle

\newpage

\section{Introduction} 
\label{sec:introduction}
 The resolved contribution induced by the interference of the electroweak operators $Q_1^q$ and $Q_{7\gamma}$ enters the inclusive $\bar B \to X_s \gamma$ decay at subleading power and constitutes one of its largest theoretical uncertainties \cite{Benzke:2020htm,Misiak:2026sqy}.

Among the resolved contributions, it is the numerically most significant. The complete resolved term  has been found to be within the range $[2.6\%, 8.9\%]$\footnote{We note that the range does not correspond to a Gaussian uncertainty. The estimate should be considered as the range within which we expect the actual values of the resolved contribution to lie, without making a statement about the most likely values within this range as noted already in Ref.~\cite{Benzke:2010js}.}. Here, the local Voloshin term of around $+3\%$ -- which was calculated based on a local expansion~\cite{Voloshin:1996gw,Ligeti:1997tc,Grant:1997ec,Buchalla:1997ky}, and which is traditionally subtracted from the resolved contribution --  was added back including  the corresponding uncertainty~\cite{Benzke:2025ekp}.

A large scale ambiguity, not included in the uncertainty above, and a strong charm-mass dependence were identified in the lowest-order result for this resolved contribution~\cite{Gunawardana:2019gep,Benzke:2020htm}. These issues motivate a systematic computation of $\alpha_s$ corrections and a consistent renormalisation-group (RG) 
resummation~\cite{Gunawardana:2019gep,Benzke:2020htm}.\footnote{The scale dependence of the LO result is not controlled. The scale of the hard functions (i.e. the Wilson coefficients) is not fixed. If one changes their scale from the hard scale to the hard-collinear scale, the final result of the resolved contribution increases by almost $40\%$. The NLO analysis will diminish this large scale ambiguity of the LO result and will replace it by well-defined scale dependence of $\mathcal{O}(\alpha_s^2)$, and may also reduce the dependence on the charm-quark mass~\cite{Gunawardana:2019gep,Benzke:2020htm}. This should lead to a significantly improved estimate of this resolved contribution and hence to a more precise Standard Model prediction for the $\bar B\to X_s\gamma$ branching ratio.}
An all-order factorisation formula in $\alpha_s$ is required to achieve this goal, separating the subleading non-local contributions into the hard functions $H_1^q$ and $H_{7\gamma}$, the hard-collinear jet function $J^{(n)}$, the anti-hard-collinear jet function $J^{(\bar n)}$, and the shape function $g_{17}$. While such a factorisation theorem at leading order was established in \cite{Benzke:2010js} and a failure of factorisation in another resolved contribution was healed by refactorisation techniques \cite{Hurth:2023paz}, we provide a formulation of the corresponding factorisation theorem that is suitable for the systematic inclusion of higher-order corrections in $\alpha_s$. 

Within this framework, we compute the hard matching functions associated with the operators $Q_1^q$ and $Q_{7\gamma}$, including the relevant mixing contributions at the required order. In addition, we review the known results for the hard-collinear jet function $J^{(n)}$, which has been computed at NLO and NNLO in~\cite{Becher:2006qw}, with the N$^3$LO result presented in~\cite{Bruser:2018rad}.

A central result of this paper is the computation of the massless anti-hard-collinear jet function at NLO. This function can be extracted from the anti-hard-collinear region of the $b\to s \gamma g$ amplitude induced by the operator $Q_1^q$. We first derive the one-loop result and subsequently evaluate the two-loop diagrams in the limit $m_c\to m_u\approx 0$. We provide the complete $u$-integrated expression, including all finite terms. While quark-mass corrections modify the finite parts, the $1/\eps$ poles are fixed by the anomalous dimensions of the hard, jet and soft functions, which do not depend on the charm-quark mass (see also~\cite{Greub:1996tg,Greub:1996wn} for the smooth massless limit of the hard penguin function). Therefore, the massless calculation captures all ultraviolet (UV) and infrared (IR) divergences required for the explicit verification of pole cancellation.

 Recently, we provided a crucial ingredient for the next-to-leading-order analysis of the resolved-photon $Q_1^q$–$Q_{7\gamma}$ interference contribution~\cite{Bartocci:2024bbf,Bartocci:2025lbl}. We computed the one-loop anomalous dimension of the corresponding soft function in the factorisation theorem, namely the subleading shape function $g_{17}$, and derived and solved its RG equation. 
 
 In addition, we found~\cite{Bartocci:2024bbf,Bartocci:2025lbl}, for inclusive as well as exclusive modes, that the anomalous dimension of such a multi-light-cone operator contains terms that are irrelevant in factorisation theorems (and therefore for physical quantities). This allows for the simplification of the renormalisation of operators in Heavy-Quark Effective Theory (HQET) that are composed of fields smeared along distinct light-cone directions. Progress has also been made in understanding the properties of such di-light-cone functions and in constructing phenomenological models for them, including higher-twist contributions~\cite{Bartocci:2026dlcda}.

 Since we have determined the divergences of all the functions entering the factorisation formula, we are able to provide an explicit proof of the cancellation of all singularities. This result constitutes a non-trivial cross-check of the aforementioned simplification of the soft functions, as it demonstrates the cancellation of all $1/\eps$ singularities at next-to-leading order between the hard, hard-collinear, anti-hard-collinear and soft contributions, including the omission of the previously identified irrelevant terms.
Furthermore, this provides a comprehensive consistency check of both the factorisation framework and our calculations. It also lays the groundwork for the inclusion of mass corrections in the anti-hard-collinear jet function at two loops and forms the basis for a future phenomenological analysis of this resolved contribution.

The manuscript is organised as follows. In Section~\ref{sec:factorisation} we derive the factorisation theorem, which is valid beyond leading order. In Section~\ref{sec:antijet_from_matching} we establish the connection between the anti-hard-collinear jet function and the $b\to s\gamma g$ amplitude. Section~\ref{sec:literature} reviews the results already available in the literature that are needed for the pole cancellation. In Section~\ref{sec:hardmatching} we compute the $\mathcal{O}(\alpha_s)$ corrections to the hard functions, and in Section~\ref{sec:antijet_function} we compute the anti-hard-collinear jet function at NLO, assuming a massless quark loop. In Section~\ref{sec:pole_cancellation} we validate the framework and the calculation by showing explicitly that all poles cancel within the factorisation formula, and in Section~\ref{sec:pole_cancellation_exclusive} we show that the same cancellation holds for the analogous exclusive decay $\bar{B}_{d,s}\to\gamma\gamma$. Finally, we conclude in Section~\ref{sec:conclusion}. In Appendix~\ref{app:setup} we summarise our conventions, including the weak effective Hamiltonian, the kinematics and power counting, and the gauge-invariant SCET building blocks. Appendix~\ref{app:distribution_definition} collects the definitions of the distributions used throughout the paper, together with useful convolution identities. In Appendix~\ref{app:u-indep-results} we give the master integrals and the diagram-by-diagram results for the $u$-integrated anti-hard-collinear jet function. 
Finally, in Appendix~\ref{consistencycheck} we verify at leading order that the two definitions of the anti-hard-collinear jet function, from the factorisation formula and from the matching condition, agree.

\section{Factorisation theorem of the resolved \texorpdfstring{$Q_1^q$--$Q_{7\gamma}$}{Q1-Q7gamma} contribution}
\label{sec:factorisation}
\begin{figure}[t]
    \centering
    \includegraphics[width=0.43\textwidth]{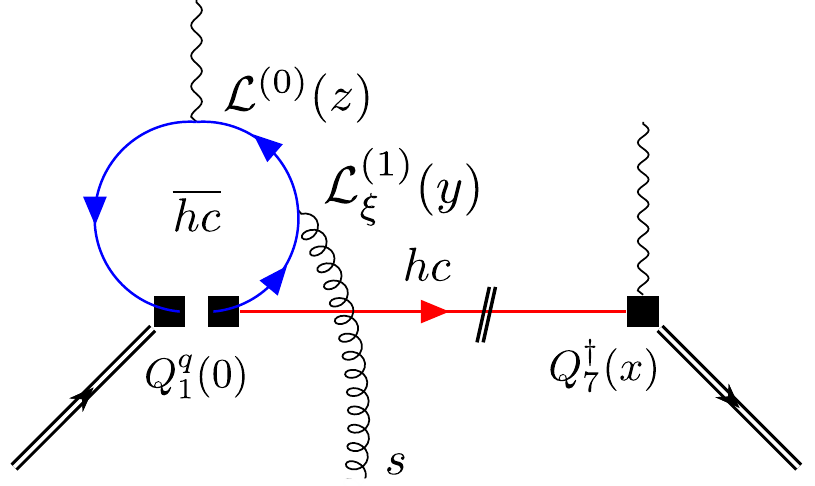}
    \caption{Schematic illustration of the factorisation theorem including  the required Lagrangian insertions (see text).}
    \label{fig:heuristic}
\end{figure}

The LO factorisation theorem has already been derived in~\cite{Benzke:2010js}. In this section, we extend the result to obtain a general factorisation theorem that is valid to compute the $\alpha_s$ corrections. Our objective is to factorise the hard, hard-collinear, anti-hard-collinear, and soft modes from the $Q_1^q$–$Q_{7\gamma}$ contribution.

Figure~\ref{fig:heuristic} provides a schematic picture to illustrate the mechanism of factorisation.\footnote{For our general conventions we refer the reader to Appendix~\ref{app:setup}.} The gauge-invariant photon field $\mathcal{A}_{\perp,\overline{hc}}$ is emitted from the anti-hard-collinear quark loop via an insertion $\mathcal{L}^{(0)}(z)$ of the leading-power QED interaction of anti-hard-collinear quarks. To compensate for the free transverse index $\mu_\perp$ in the anti-hard-collinear jet function, a subleading insertion $\mathcal{L}^{(1)}_\xi(y)$ is required, where $y_+ \equiv r \bar{n}$ and $x_- \equiv t n$. This insertion generates a soft shape function that includes an additional soft gluon field. The red and blue lines in the figure represent the soft Wilson line structure in the $n^\mu$ and $\bar{n}^\mu$ directions, respectively. Specifically, the subleading Lagrangian in terms of decoupled fields takes the form:
\begin{equation}
    \mathcal{L}_\xi^{(1)}(y)=\bar{\chi}_{\overline{hc}}(y)i y^\mu_\perp S_{\bar{n}}^\dagger(y_+)\left(-ig_s\bar{n}^\nu G_{\mu\nu}(y_+)\right)S_{\bar{n}}(y_+)\frac{\nslash}{2}{\chi}_{\overline{hc}}(y),
\end{equation}
where $S_{\bar{n}}$ and  $S_{\bar{n}}^\dagger$ denote the soft Wilson lines from the decoupling of the anti-hard-collinear fermion fields.

The first step in deriving the factorisation theorem is the hard matching. In this context, the electroweak operators $Q_1^q$ (with $q=c,u$) and $Q_{7\gamma}$ must be matched onto 
soft-collinear effective theory (SCET).
In QCD one has the two local electroweak operators:\footnote{In this section we use the Fierz-transformed version of the regular $Q_1$. It is well-known that the renormalisation properties of the operator change starting at order $\alpha_s$, but this computational issue is well under control (see~\cite{Aebischer:2023djt} and references therein).}
\begin{equation}
\label{eq:Q1-def}
Q_1^q=[\bar{s}_j \gamma^\mu(1-\gamma_5)b_i][\bar{q}_i\gamma_\mu(1-\gamma_5)q_j], \,\,\,\,\,\,\,\, Q_{7\gamma}=-\frac{e}{8\pi^2}m_b\bar{s}\sigma_{\mu\nu}(1+\gamma_5)F^{\mu\nu}b\,.
\end{equation}
The explicit indices $i$ and $j$ represent colour indices, while Lorentz indices are suppressed in this notation.

 The operator $Q_1^q$ corresponds to a (non-local) $B$-type operator in SCET, while $Q_{7\gamma}$ is matched onto a (local) $A$-type current operator (see~\cite{Beneke:2017ztn} for the precise classification of these current types). The gauge-invariant building blocks are used for the gauge fields and the fermionic fields within SCET (see~\cite{Hill:2002vw}, also  Appendix~\ref{app:setup}). QCD operators are written as $Q_i$, while SCET operators are denoted by ${\cal O}_i$.

The electroweak operator $Q_{7\gamma}$ with its Wilson coefficient $C_{7\gamma}$ is matched onto 
\begin{equation}
    \label{eq:O7-matching}
H_{7\gamma}\,   {\cal O}_{7 \gamma}(x) = H_{7\gamma} \frac{-m_b}{4\pi^2}e^{-im_bv\cdot x}\bar{\chi}_{hc}(x)\frac{\slashed{\bar{n}}}{2}[in\cdot\partial \slashed{\mathcal{A}}_{\perp,\overline{hc}}(x)](1+\gamma_5)S_n^\dagger(x_-)h_v(x_-)\,.
\end{equation}
The SCET operator ${\cal O}_{7 \gamma}$ is already given in terms of decoupled fields. It is defined with gauge-invariant fermionic fields and the gauge-invariant transverse photon field, which is given as
\begin{equation*}
{\mathcal{A}}_{\perp,\overline{hc}}^{\nu}(x)=
e\left(A_{\perp}^{\nu}-\frac{\partial_{\perp}^{\nu}}{n\cdot\partial}\, n\cdot A\right).\end{equation*}
The matching equation also involves a multiplicative hard matching coefficient of the general form 
$H_{7 \gamma} = C_{7\gamma} ( 1 + {\cal O}(\alpha_s))$.

For the concrete matching, we can calculate the QCD operator matrix element $b\left(p_b\right)\to s\left(p_s\right) \gamma(q)$, where the momentum $p_b$ is hard, $p_s$ is hard-collinear and $q$ is anti-hard-collinear. 
The matching condition is given by\footnote{\label{fn:mixing}Additional mixing contributions from other electroweak operators contribute to this matching equation. These mixing contributions are proportional to the tree-level matrix element of $Q_{7\gamma}$. We suppress them in our notation in this section, but they will be discussed in Section~\ref{sec:hardmatching}. We anticipate that the general structure is $H_{7\gamma} = C_{7\gamma} (1 + \alpha_s h_{77}) +  C_{8g} \alpha_s h_{78} + C_1^q \alpha_s h_{71} + {\cal O}(\alpha_s^2)$. The same comment applies to the matching equation~\eqref{eq:matching1}: $H_1^q = C_1^q (1 + \alpha_s k_{11}) + C_2^q \alpha_s k_{12}$.} 
\begin{equation}
\label{eq:matching7}
C_{7 \gamma}\, \left\langle Q_{7\gamma} \right\rangle = H_{7\gamma}  \,   \left\langle {\cal O}_{7\gamma} \right\rangle \,,
\end{equation}
which fixes the hard matching function $H_{7\gamma}$. The brackets $\left\langle \,\, \right\rangle$ indicate that the matrix elements of the operators are considered.

Similarly, the electroweak operator  $Q_1^q$ with its Wilson coefficient $C_1^q$ is matched onto   
\begin{align}  \label{matchO12}
  & \int d\tilde{s} \Tilde{H}_1^q(s) {\cal O}_1^q(s,x) =\\  
&\int d\tilde{s} \Tilde{H}_1^q(s)e^{-im_b v \cdot x} (\bar{\chi}_{hc}(x)S_n^\dagger(x_-))^j\slashed{\bar{n}}(1-\gamma_5)h_v^i(x_-) 
 ({\bar{\chi}}_{\overline{hc}}(x+sn)S^{\dagger}_{\bar{n}}(x_+))^i \frac{\slashed{n}}{2}(1-\gamma_5)  (S_{\bar{n}}(x_+)\chi_{\overline{hc}}(x))^j\,,  \nonumber
   \end{align}
where the SCET operator ${\cal O}_1^q$ is again given in terms of decoupled fields. Here $\tilde{s} = s\, n\cdot q$, and $q$ is the anti-hard-collinear momentum of the photon.  
After separating the soft Wilson lines we end up with 
\begin{equation}
 \int d\tilde{s} \Tilde{H}_1^q(s)e^{-im_b v \cdot x}\bar{\chi}_{hc}^{\hat{j}}(x)\slashed{\bar{n}}(1-\gamma_5)h_v^i(x_-)\bar{\chi}_{\overline{hc}}^{\hat{i}}(x+sn)\frac{\slashed{n}}{2}(1-\gamma_5)\chi_{\overline{hc}}^{\Tilde{j}}(x)S_{n}^{\dagger\hat{j}j}(x_-)S_{\bar{n}}^{\dagger\hat{i}i}(x_+)S_{\bar{n}}^{j\Tilde{j}}(x_+) .
\end{equation}
To fix the hard function $\Tilde{H}_1^q$, we can consider the matching condition via the amplitude $b(p_b) \to s(p_s) + q(p_q) +\bar{q}(p_{\bar{q}})$ in QCD and SCET, where the momentum $p_b$ is hard, $p_s$ is hard-collinear, $p_q$ and $p_{\bar{q}}$ are anti-hard-collinear: 
\begin{equation}
\label{eq:matching1}
C_1^q\, \left\langle Q_1^q \right\rangle =  \int d\tilde{s} \Tilde{H}_1^q(s) \left\langle{\cal O}_1^q(s,x) \right\rangle\,.
\end{equation}

The differential decay rate is related to a restricted discontinuity of the forward $B$-meson matrix element of the product of two weak effective Hamiltonians:
\begin{equation}\label{THH}
   d\Gamma(\bar B\to X_s\gamma) = \frac{1}{2M_B i} \mbox{Disc}_{\rm \,restr.}\,
   \Big[ i\int d^4 x\,\langle\bar B| {\cal H}_{\rm eff}^\dagger(x)\,
   {\cal H}_{\rm eff}(0) |\bar B\rangle \Big] \,.
\end{equation}
This discontinuity is restricted by requiring that the cut propagators must include a photon and a strange quark.
The contribution of interest is then expressed in terms of the time-ordered product:
\begin{align}
      d\Gamma(\bar B\to X_s\gamma) =&\, \frac{\mathcal{N}_H}{2M_B i}\, \mbox{Disc}_{\rm \,restr.}\,
   \Big[ i\int d^4x\, \langle\bar B|T\{{C_1^q}\,Q_1^q(0),{C_{7\gamma}^*}\,Q_{7\gamma}^\dagger(x), \nonumber \\
   &\qquad\qquad\qquad\qquad\qquad\qquad i\int d^4y\, \mathcal{L}^{(1)}_\xi(y),\, i\int d^4z\,  \mathcal{L}^{(0)}(z)\}|\bar B\rangle \Big]\,,
\end{align}
{where the CKM factors and the Fermi constant of the effective Hamiltonian are collected in
\begin{equation}
\label{eq:Nweak}
\mathcal{N}_H\equiv\Big(\frac{G_F}{\sqrt{2}}\Big)^2\lambda_q\Big(\sum_{q'=u,c}\lambda_{q'}\Big)^*=-\frac{G_F^2}{2}\,\lambda_q\lambda_t^*\,,
\end{equation}
using CKM unitarity.} {The Wilson coefficients $C_1^q$ and $C_{7\gamma}^*$ are absorbed into $\tilde H_1^q$ and $H_{7\gamma}^*$ through the matching conditions~\eqref{eq:matching1} and~\eqref{eq:matching7}.}
Here $\mathcal{L}^{(0)}(z)$ denotes the leading-power anti-hard-collinear QED interaction, which describes the emission of the photon from the anti-hard-collinear quark loop.
By inserting the explicit definitions, we obtain\footnote{Here and in the following, $(in\cdot\partial)^*\equiv -in\cdot\partial$ arises from the Hermitian conjugate of ${\cal O}_{7\gamma}$ and acts only on the photon field.}
\begin{align}
          d\Gamma(\bar B\to X_s\gamma) =&  \frac{{\mathcal{N}_H}}{2M_B i} \mbox{Disc}_{\rm \,restr.}\,
   \Big[ i \int d^4x \langle\bar B|T\{ \\
   &\times\int d\tilde{s}\, \Tilde{H}_1^q(s) \bar{\chi}_{hc}^{\hat{j}}(0)\slashed{\bar{n}}(1-\gamma_5)h_v^i(0)\bar{\chi}_{\overline{hc}}^{\hat{i},(u)}(sn)\frac{\slashed{n}}{2}(1-\gamma_5)\chi_{\overline{hc}}^{\Tilde{j},(1-u)}(0)S_{n}^{\dagger\hat{j}j}(0)S_{\bar{n}}^{\dagger\hat{i}i}(0)S_{\bar{n}}^{j\Tilde{j}}(0) \nonumber\\
   &\times H_{7\gamma}^*\frac{-m_b}{4\pi^2}e^{im_bv\cdot x}\bar{h}_v(x_-)S_n(x_-)(1-\gamma_5)[(in\cdot\partial)^*\slashed{\mathcal{A}}_{\perp,\overline{hc}}(x)]\frac{\slashed{\bar{n}}}{2}\, \chi_{hc}(x) \nonumber\\
   &\times i\int d^4y \bar{\chi}_{\overline{hc}}(y)i y^\mu_\perp S_{\bar{n}}^\dagger(y_+)\left(-ig_s\bar{n}^\nu G_{\mu\nu}(y_+)\right)S_{\bar{n}}(y_+)\frac{\nslash}{2}{\chi}_{\overline{hc}}(y), i\int d^4z \mathcal{L}^{(0)}(z)\,\,\,\}|\bar B\rangle \Big]. \nonumber
\end{align}
 
Since the states factorise, the matrix elements can be evaluated separately in the hard-collinear, anti-hard-collinear and soft sectors.

We consider first the fermionic currents containing hard-collinear and soft fields in the second and third lines:
\begin{equation}
\bar{h}_v(x_-)S_n(x_-)(1-\gamma_5)[(in\cdot\partial)^*\slashed{\mathcal{A}}_{\perp,\overline{hc}}(x)] \frac{\slashed{\bar{n}}}{2}\,\chi_{hc}(x)  \,\,\, \,\,\,\bar{\chi}_{hc}(0)\slashed{\bar{n}}(1-\gamma_5)h_v(0)\,.
\end{equation}
The hard-collinear fields can already be fully separated at this stage by using the momentum-space representation of the hard-collinear jet function,
\begin{equation}
        {\langle 0|T\{[\bar{\chi}^{i}_{hc,o}(0), \chi^{j}_{hc,p}(x)]\}| 0\rangle \equiv (-1)\,\delta^{ji}\int\frac{d^4p}{(2\pi)^4}\, e^{-ip\cdot x} \Big(\frac{\slashed{n}}{2}\Big)_{po} \bar{n}\cdot p \, {\cal J}^{(n)}(p^2;\mu)\,,}
  \label{jetmomentum}  \end{equation}
where $o$ and $p$ denote Dirac indices. It then follows that
\begin{equation}
    \langle 0|T\{[\bar{\chi}_{hc,o}^{\hat{j}}(0),\frac{\slashed{\bar{n}}}{2}\chi_{hc,p}^l(x)]\}| 0\rangle \propto  (\frac{\slashed{\bar{n}} \slashed{n}}{4})_{po}\,. 
\end{equation}
This matrix, the hard-collinear projector, acts trivially on the fermionic string between the soft fields, owing to the relation $(\frac{\slashed{\bar{n}} \slashed{n}}{4}) \slashed{\bar{n}} = \slashed{\bar{n}}$. In the separated hard-collinear matrix element, on the other hand, the Dirac indices are now contracted, which reduces it to the scalar part of the hard-collinear jet function in momentum space (see~\eqref{jetfinal}). We thus arrive at
\begin{align}
        d\Gamma(\bar B\to X_s\gamma) = & \frac{{\mathcal{N}_H}}{2M_B i}  {\color{black} (-2)} (H_{7\gamma}^*\frac{-m_b}{4\pi^2})\mbox{Disc}_{\rm \,restr.}\,
   \Big[ i \int d^4x e^{im_bv\cdot x} \int d\tilde{s}\, \Tilde{H}_1^q(s) i\int d^4y \label{rearrange} \\
   &\times\frac{1}{2}\langle 0|T\{[\bar{\chi}_{hc,o}^{\hat{j}}(0),\frac{\slashed{\bar{n}}}{2}\chi_{hc,o}^l(x)]\}| 0\rangle \nonumber\\
   &\times \langle \bar{B}|T\{[(\bar{h}_v(x_-)S_n(x_-))^l[(in\cdot\partial)^*\slashed{\mathcal{A}}_{\perp,\overline{hc}}(x)]{\color{black} \slashed{\bar{n}} (1 - \gamma_5)}h_v^i(0)] \nonumber\\
   &\times[\bar{\chi}_{\overline{hc}}^{\hat{i},(u)}(sn)\frac{\slashed{n}}{2}(1-\gamma_5)\chi_{\overline{hc}}^{\Tilde{j},(1-u)}(0)]S_{n}^{\dagger\hat{j}j}(0)S_{\bar{n}}^{\dagger\hat{i}i}(0)S_{\bar{n}}^{j\Tilde{j}}(0) \nonumber\\
   &\times [\bar{\chi}_{\overline{hc}}^{k}(y)i y^\mu_\perp \big(S_{\bar{n}}^\dagger(y_+)\left(-ig_s\bar{n}^\nu G_{\mu\nu}(y_+)\right)S_{\bar{n}}(y_+)\big)^{k\hat k}\frac{\nslash}{2}{\chi}_{\overline{hc}}^{\hat k}(y)], i\int d^4z \mathcal{L}^{(0)}(z)\}|\bar B\rangle\Big]\,. \nonumber
\end{align}
The additional factor $(-2)$ arises from the anti-commutation of the hard-collinear fields and from the relation $(1 - \gamma_5)(1-\gamma_5)=2 (1-\gamma_5)$. For the colour structure of the hard-collinear matrix element we have
\begin{equation}
    \langle 0|T\{[\bar{\chi}_{hc,o}^{\hat{j}}(0),\tfrac{\slashed{\bar{n}}}{2}\chi_{hc,o}^l(x)]\}| 0\rangle=\delta^{l\hat{j}}\, \langle 0|T\{[\bar{\chi}_{hc,o}(0),\tfrac{\slashed{\bar{n}}}{2}\chi_{hc,o}(x)]\}| 0\rangle\,.
\end{equation}
In the following, hard-collinear matrix elements without explicit colour indices are understood in this sense, see~\eqref{jetfinal}.

Grouping the anti-hard-collinear and soft fields, we note that the contraction of the colour indices in the soft matrix element implies $\delta_{\Tilde{j}k}\delta_{\hat{i}\hat{k}}$, which fixes the colour structure of the anti-hard-collinear function; we relabel $\hat i\to i$ and $\tilde j\to j$ in the following. Moreover, the anti-hard-collinear and soft functions are proportional to $g_\perp^{\mu\alpha}$ and $g^\perp_{\mu\alpha}$, respectively. Putting everything together, we find
\begin{align}
        d\Gamma(\bar B\to X_s\gamma) =& \frac{{\mathcal{N}_H}}{2M_B i}       H_{7\gamma}^*\, (-2)\,         \frac{-m_b}{4\pi^2}\mbox{Disc}_{\rm \,restr.}\,
   \Big[ i\int d^4x e^{im_bv\cdot x} \int d\tilde{s}\, \Tilde{H}_1^q(s) i\int d^4y \\
   &\times\frac{1}{2}\langle 0|T\{[\bar{\chi}_{hc,o}(0),\frac{\slashed{\bar{n}}}{2}\chi_{hc,o}(x)]\}| 0\rangle \nonumber\\
   &\times\langle 0|T\{[\bar{\chi}_{\overline{hc}}^{i,(u)}(sn)\frac{\slashed{n}}{2} { (1 - \gamma_5)}    \chi_{\overline{hc}}^{j,(1-u)}(0)],[\bar{\chi}_{\overline{hc}}^{j}(y)iy^\mu_\perp\frac{\slashed{n}}{2}\chi_{\overline{hc}}^{i}(y)], \nonumber\\
   &\qquad i\int d^4z \mathcal{L}^{(0)}(z)\}[(in\cdot\partial)^*{\mathcal{A}}_\mu^{\perp,\overline{hc}}(x)]|0\rangle \nonumber\\
   &\times\langle \bar{B}|(\bar{h}_vS_n)(x_-) (S_n^\dagger S_{\bar{n}})(0){ (-i)}\gamma^\alpha_\perp \bar{n}^\beta(S_{\bar{n}}^\dagger g_s G_{\alpha\beta}S_{\bar{n}})(y_+) { \slashed{\bar{n}}(1-\gamma_5)}(S_{\bar{n}}^\dagger h_v)(0)|\bar B\rangle\Big]\,. \nonumber
\end{align}

We now split the anti-hard-collinear matrix element into the photon propagator, which is cut, and the anti-hard-collinear jet function, which is defined at the amplitude level:
\begin{align}
      d\Gamma(\bar B\to X_s\gamma) =& \frac{{\mathcal{N}_H}}{2M_B i}      H_{7\gamma}^*\, (-2)\,    \frac{-m_b}{4\pi^2}\mbox{Disc}_{\rm \,restr.}\,
   \Big[ i\int d^4x e^{im_bv\cdot x} \int d\tilde{s}\, \Tilde{H}_1^q(s)\, i\int d^4y\, i\int d^4z \\
   &\times\frac{1}{2}\langle 0|T\{[\bar{\chi}_{hc,o}(0),\frac{\slashed{\bar{n}}}{2}\chi_{hc,o}(x)]\}| 0\rangle \nonumber\\
   &\times\langle 0|T\{\mathcal{A}_\lambda^{\perp,\overline{hc}}(z)(in\cdot\partial)^*{\mathcal{A}}_\mu^{\perp,\overline{hc}}(x)\}|0\rangle \nonumber\\
   &\times\langle 0|T\{[\bar{\chi}_{\overline{hc}}^{i,(u)}(sn)\frac{\slashed{n}}{2}  {(1 - \gamma_5)}     \chi_{\overline{hc}}^{j,(1-u)}(0)],[\bar{\chi}_{\overline{hc}}^{j}(y)iy^\mu_\perp\frac{\slashed{n}}{2}\chi_{\overline{hc}}^{i}(y)], \nonumber\\
&\times[\bar{\chi}_{\overline{hc}}(z)i \overleftarrow{\not\partial_\perp}\frac{1}{{ i}n\cdot \overleftarrow{\partial}+i0}\gamma_\perp^\lambda   \frac{\slashed{n}}{2}\chi_{\overline{hc}}(z) \,\, + \,\,\bar{\chi}_{\overline{hc}}(z)\gamma_\perp^\lambda \frac{1}{{ i}n\cdot \overrightarrow{\partial}+i0}  i \overrightarrow{\not\partial_\perp} \frac{\slashed{n}}{2}        \chi_{\overline{hc}}(z)]\}|0\rangle \nonumber\\
   &\times\langle \bar{B}|(\bar{h}_vS_n)(x_-) (S_n^\dagger S_{\bar{n}})(0){ (-i)}\gamma^\alpha_\perp\bar{n}^\beta(S_{\bar{n}}^\dagger g_s G_{\alpha\beta}S_{\bar{n}})(y_+)({\slashed{\bar{n}}(1-\gamma_5)}  S_{\bar{n}}^\dagger h_v)(0)|\bar B\rangle\Big]\,. \nonumber
\end{align}
Since the photon propagator satisfies
\begin{equation}
\label{eq:photon_gperp}
    \langle0| T\{ \mathcal{A}_\lambda^{\perp,\overline{hc}}(z)(in\cdot\partial)^*{\mathcal{A}}_\mu^{\perp,\overline{hc}}(x)\}|0\rangle \propto ~ g_{\lambda\mu}^\perp\,, 
\end{equation}
the indices $\lambda$ and $\mu$ can also be contracted in the anti-hard-collinear jet function.

The final step is to express the various matrix elements as momentum-space distributions:
\begin{itemize}
     \item We define the hard-collinear jet function ${\cal J}^{(n)}$ in momentum space as
          \begin{equation}
        \frac{1}{2} \langle 0|T\{[\bar{\chi}^{i}_{hc,o}(0),\frac{\slashed{\bar{n}}}{2}\chi^{j}_{hc,o}(x)]\}| 0\rangle \equiv (-1)\,\delta^{ji}\int\frac{d^4p}{(2\pi)^4}\, e^{-ip\cdot x}\,  \bar{n}\cdot p \, {\cal J}^{(n)}(p^2;\mu)\,,
\label{jetfinal}    \end{equation}
    where $p$ is the hard-collinear momentum flowing into the point $x$. Here we have again used the standard convention for the jet function introduced in~\eqref{jetmomentum}.
    \item We define the shape function $g_{17}$ in momentum space as
    \begin{align}
        &\langle \bar{B}|(\bar{h}_vS_n)(x_-)\slashed{\bar{n}}(1+\gamma_5)(S_n^\dagger S_{\bar{n}})(0)i\gamma^\alpha_\perp\bar{n}^\beta(S_{\bar{n}}^\dagger g_s G_{\alpha\beta}S_{\bar{n}})(y_+)(S_{\bar{n}}^\dagger h_v)(0)|\bar B\rangle \label{shapedefinition} \\
        &\equiv 2M_B\int d\omega e^{i\omega t}\int d\omega_1 e^{i\omega_1 r}g_{17}(\omega,\omega_1)\,, \nonumber
\end{align}
    with $x_-=tn$ and $y_+=r\bar{n}$.
    \item We define the photon propagator $J_\gamma$ as
    \begin{equation}
       i \int d^4x e^{+i q\cdot(x-z)} \langle0| T\{(in\cdot\partial)^*{\mathcal{A}}_\mu^{\perp,\overline{hc}}(x),\mathcal{A}_\lambda^{\perp,\overline{hc}}(z)\}|0\rangle= i g_{\lambda\mu}^\perp {n}\cdot q \ e^2 \, J_\gamma(q^2),
    \end{equation}
    where $q$ is the anti-hard-collinear momentum flowing into the point $x$. Working at leading order in the electromagnetic coupling, we have $J_\gamma = i /(q^2 + i0)$, and its discontinuity reads
\begin{equation}
    \mbox{Disc}[ i \int d^4x e^{+i q\cdot(x-z)} \langle0| T\{(in\cdot\partial)^*{\mathcal{A}}_\mu^{\perp,\overline{hc}}(x),\mathcal{A}_\lambda^{\perp,\overline{hc}}(z)\}|0\rangle] = 2\pi i 
g_{\lambda\mu}^\perp {n}\cdot q  e^2 \, \delta(q^2) \theta(q^0)\,.
   \end{equation}  
   It follows that
   \begin{equation}
       \langle0| T\{(in\cdot\partial)^*{\mathcal{A}}_\mu^{\perp,\overline{hc}}(x),\mathcal{A}_\lambda^{\perp,\overline{hc}}(z)\}|0\rangle = \int\frac{d^4q}{(2\pi)^4 }  e^{- i q\cdot(x-z)}g_{\lambda\mu}^\perp\,  {n}\cdot q \,  e^2\ J_\gamma(q^2),
    \end{equation}
    where, as above, $q$ is the anti-hard-collinear momentum flowing into the point $x$.
    
    \item We define the anti-hard-collinear jet function $J^{(\bar{n})}$ in momentum space as
    \begin{align}
        J^{(\bar{n})}(u,\omega_1) =& \int \frac{d\tilde{s}}{2\pi}\, e^{-i\tilde{s}u}\int d^4y e^{i\frac{\omega_1}{2}n\cdot y} 
         \int d^4z e^{i q\cdot z} \label{antijet1} \\
        &\times \langle 0|T\{[\bar{\chi}_{\overline{hc}}^{i,(u)}(sn)\frac{\slashed{n}}{2} {(1 - \gamma_5)}     \chi_{\overline{hc}}^{j,(1-u)}(0)],[\bar{\chi}_{\overline{hc}}^{j}(y)iy^\mu_\perp\frac{\slashed{n}}{2}\chi_{\overline{hc}}^{i}(y)], \nonumber\\
   &\times[\bar{\chi}_{\overline{hc}}(z)i \overleftarrow{\not\partial_\perp}\frac{1}{{ i}n\cdot \overleftarrow{\partial}+i0}\gamma_\perp^\lambda   \frac{\slashed{n}}{2}\chi_{\overline{hc}}(z) \,\, + \,\,\bar{\chi}_{\overline{hc}}(z)\gamma_\perp^\lambda \frac{1}{{ i}n\cdot \overrightarrow{\partial}+i0}  i \overrightarrow{\not\partial_\perp} \frac{\slashed{n}}{2}        \chi_{\overline{hc}}(z)]\}|0\rangle \,.  \nonumber
\end{align}
    Here the indices $\lambda$ and $\mu$ are understood to be contracted with $g^\perp_{\lambda\mu}$, see~\eqref{eq:photon_gperp}.
\item Finally, the hard function $H_1^q(u)$ is defined through
\begin{equation} \label{hardp}
    \Tilde{H}_1^q(s)\equiv\int\frac{du}{2\pi}e^{-ius(n\cdot q)}H_1^q(u) ,
\end{equation}
where $q$ is the anti-hard-collinear photon momentum.
\end{itemize}
Using these definitions, together with the fact that the decay rate is isotropic and depends only on $E_\gamma$ ($d^3 q=d|q| |q|^2 d\Omega = E_\gamma^2 d E_\gamma 4\pi$), we arrive at the factorisation formula 
\begin{align}
        d\Gamma(\bar B\to X_s\gamma) =&    {\mathcal{N}_H} \frac{e^2Q_qm_b}{4\pi^4}H_{7\gamma}^*   E_\gamma^2 d E_\gamma\int du H_1^q(u)\int d\omega_1 \, J^{(\bar{n})}(u,\omega_1) \label{FactorisationFormula} \\
       &\times\int d\omega\, g_{17}(\omega,\omega_1)\,
\frac{1}{i}\,\mbox{Disc}\Big\{ i\, m_b\, {\cal J}^{(n)}\big(m_b(m_b-2E_\gamma+\omega)\big)\Big\}. \nonumber
\end{align}

This expression corresponds to the diagrams in which the operator $Q_1^q$ belongs to the amplitude and $Q_{7\gamma}$ to the complex-conjugate amplitude. The mirror diagrams, in which the two weak vertices are interchanged, yield the complex conjugate of this expression~\cite{Benzke:2010js}. The sum of both contributions is therefore twice the real part of the expression above.

Furthermore, the resulting rate contains two types of complex phases. Weak (CP-violating) phases enter through the CKM factors collected in $\mathcal{N}_H$ and, beyond the Standard Model, through the Wilson coefficients. Strong (CP-conserving) phases arise from the anti-hard-collinear jet function $J^{(\bar n)}$, which is defined in terms of full rather than cut propagators and is therefore complex, and, beyond leading order, from the imaginary parts of the hard matching coefficients. The shape function $g_{17}$ is real~\cite{Benzke:2010js}. Under CP conjugation the weak phases change sign while the strong phases remain unchanged. Writing the contribution of the $\bar B$ decay as $2\,\mathrm{Re}[w\,s]$, with $w$ ($s$) collecting the weak (strong) phases, the corresponding $B$-meson contribution reads $2\,\mathrm{Re}[w^*s]$, and the CP average separates the two types of phases, $\mathrm{Re}[w\,s]+\mathrm{Re}[w^*s]=2\,\mathrm{Re}\,w\;\mathrm{Re}\,s$. Using these results, we arrive at the factorisation formula for the CP-averaged photon energy spectrum,\footnote{As noted above, we suppress mixing contributions in the notation of this section; according to footnote~\ref{fn:mixing}, combinations of Wilson coefficients other than $C_1^q$ and $C_{7\gamma}$ enter the hard functions in an obvious way.} 
\begin{align}
        \frac{d\Gamma(\bar B\to X_s\gamma)}{dE_\gamma}\bigg|_{\rm CP\,avg.} =&\ \mathrm{Re}\big[\mathcal{N}_H\, C_1^q C_{7\gamma}^*\big]\;2 \;\frac{e^2Q_qm_b}{4\pi^4}\; E_\gamma^2\;
        \mathrm{Re}\Bigg[\int du\, \frac{H_1^q(u)}{C_1^q}\,\frac{H^*_{7\gamma}}{C^*_{7\gamma}}\int d\omega_1 \, J^{(\bar{n})}(u,\omega_1) \label{FactorisationFormulaCP} \\
       &\times\int d\omega\, g_{17}(\omega,\omega_1)\,
\frac{1}{i}\,\mbox{Disc}\Big\{ i\, m_b\, {\cal J}^{(n)}\big(m_b(m_b-2E_\gamma+\omega)\big)\Big\}\Bigg]\,. \nonumber
\end{align}
At leading order, Equation~\eqref{FactorisationFormulaCP} reproduces the result given in Equation (42) of~\cite{Benzke:2010js}.

In the following sections we collect and compute all gauge-invariant functions entering the factorisation theorem at NLO, in order to verify the cancellation of poles.

\section{Anti-hard-collinear jet function via matching}
\label{sec:antijet_from_matching}
In this section we derive the explicit matching condition for the anti-hard-collinear jet function, which is defined at the amplitude level. This allows us to extract the function uniquely from the $b \to s \gamma\,  g$ amplitude~\cite{Boer:2023yde}.

We start from the two QCD four-quark operators $Q^{q}_1$ and $Q^{q}_2$, which we use in the calculation of the $b \to s \gamma\,  g$ amplitude in the following sections. Following the conventions of~\cite{Beneke:2001ev}, this basis consists of the singlet and octet operators

 \begin{equation}
 \label{eq:Q1-Q2}
 Q_1^q = {\bar s}_j \gamma^\mu (1- \gamma_5)  q_j\,\,\,  {\bar q}_i \gamma_\mu (1-\gamma_5)  b_i \,, \qquad \qquad   Q_2^q = \bar s_j  \gamma^\mu (1 - \gamma_5)  q_i \,\,\, {\bar q}_i  \gamma_\mu (1 -\gamma_5)   b_j\,,  
\end{equation}
where, as before, the explicit indices are colour indices and Lorentz indices are suppressed. The operators $Q_1^{q}$ and $Q_2^{q}$ are matched onto the SCET operators
   \begin{align}
   \label{eq:O1-matching}
        \begin{split}
 {\cal O}_1^{q}(s) =     \,\, &  \bar{\chi}_{hc}^j(0)   \gamma_\mu^\perp(1-\gamma_5)  \chi_{\overline{hc}}^{j}(0)      \bar{\chi}_{\overline{hc}}^{i}(sn) \gamma^\mu_\perp (1-\gamma_5)     h_v^i(0)  \,,\\
 {\cal O}_2^{q}(s)=  \,\,   & \bar{\chi}_{hc}^j(0)   \gamma_\mu^\perp(1-\gamma_5)  \chi_{\overline{hc}}^{i}(0)      \bar{\chi}_{\overline{hc}}^{i}(sn) \gamma^\mu_\perp (1-\gamma_5)     h_v^j(0)  \,.
    \end{split}
\end{align}

Factorisation of the $b \to s \gamma\, g$ amplitude requires the insertion of the subleading SCET Lagrangian, which produces the physical soft gluon, and of the leading-power anti-hard-collinear QED interaction, which produces the photon. We therefore consider the SCET matrix element
\begin{equation}
        \langle s \gamma g| T\{ {\cal  O}_{1,2}^{q}(s),\, i\int d^4y\,  \mathcal{L}^{(1)}_{\xi} (y),\, i\int d^4z\,  \mathcal{L}^{(0)} (z)\}|b\rangle\,.
\end{equation}
The operators ${\cal O}_{1,2}^q$ have mass dimension $6$, and the time-ordered product with the subleading Lagrangian scales as $\lambda^{7}$ (with $\lambda^2=\Lambda_{\rm QCD}/m_b$, see Appendix~\ref{app:setup}). The external on-shell particles are described by the SCET fields $h_v$, ${\mathcal{A}}^\gamma_{\perp,\overline{hc}}$, ${\mathcal{A}}_{\perp,{s}}$ and $\chi_{hc}$. We therefore need boost-invariant SCET operators with the same mass dimension and power counting. Concretely, this requires a derivative acting on a large momentum component, $i n\cdot\partial$ or $i \bar{n}\cdot\partial$, to match the mass dimension, and a factor $\frac{\slashed{\bar{n}}}{2}$ to ensure boost invariance. This leads to the two SCET operators
\begin{align}
\begin{split}
\bar{\chi}_{hc}(0) \frac{\slashed{\bar{n}}}{2} (i n\cdot\partial \slashed{\mathcal{A}}^\gamma_{\perp,\overline{hc}}(0)) \slashed{\mathcal{A}}_{\perp,{s}}(r\bar{n}) h_v(0) \equiv {\cal O}^{a}(r)\,,\\
\bar{\chi}_{hc}(0) \frac{\slashed{\bar{n}}}{2} \slashed{\mathcal{A}}_{\perp,{s}}(r\bar{n}) (i n\cdot\partial \slashed{\mathcal{A}}^\gamma_{\perp,\overline{hc}}(0)) h_v(0) \equiv {\cal O}^{b}(r)\,.
\end{split} 
\end{align}

We can now define the anti-hard-collinear coefficient functions through the matching condition
\begin{align}
        \bra{s\gamma g}&T\{\tilde{\cal O}_{1,2}^q(u),i\int d^4 y\, \mathcal{L}_{\xi}^{(1)}(y),i\int d^4 z\, \mathcal{L}^{(0)}(z)\}\ket{b}\equiv \label{antijetmatch} \\
    &g_sQ_q\int\frac{dr}{2\pi}e^{-ir\bar{n}\cdot k_g}\Bigg(J^{(\bar{n}),a}(u,r)\bar{u}_{\xi}^{(s)}\frac{\nbslash}{2}(n\cdot k_\gamma \slashed{\epsilon}_\perp^\gamma)\slashed{\epsilon}_{s,\perp}h_v+J^{(\bar{n}),b}(u,r)\bar{u}_{\xi}^{(s)}\frac{\nbslash}{2}\slashed{\epsilon}_{s,\perp}(n\cdot k_\gamma \slashed{\epsilon}_\perp^\gamma)h_v\Bigg)= \nonumber \\
    &g_sQ_q\Bigg(J^{(\bar{n}),a}(u,\bar{n}\cdot k_g)\bar{u}_{\xi}^{(s)}\frac{\nbslash}{2}(n\cdot k_\gamma \slashed{\epsilon}_\perp^\gamma)\slashed{\epsilon}_{s,\perp}h_v+J^{(\bar{n}),b}(u,\bar{n}\cdot k_g)\bar{u}_{\xi}^{(s)}\frac{\nbslash}{2}\slashed{\epsilon}_{s,\perp}(n\cdot k_\gamma \slashed{\epsilon}_\perp^\gamma)h_v\Bigg)\,, \nonumber
\end{align}
where $\tilde{\cal O}^q_{1,2}(u) =  \int \frac{d\tilde{s}}{2\pi} \,\,  e^{- i u s n\cdot  q} {\cal O}^q_{1,2}(s)$, in accordance with~\eqref{matchO12} and~\eqref{hardp}. This matching condition holds to all orders in $\alpha_s$.

We now analyse how the two matching coefficients $J^{(\bar{n}),a}$ and $J^{(\bar{n}),b}$ enter the decay rate. Starting from~\eqref{THH}, we combine the operator $Q_{7\gamma}^\dagger$, given in~\eqref{eq:Q1-def}, with the right-hand side of the matching condition. The matrix element of $Q_{7\gamma}^\dagger$ is, up to a constant prefactor,
\begin{equation}
\langle b|Q_{7\gamma}^\dagger|s\gamma\rangle\propto\bar{h}_v(1-\gamma_5)\,n\cdot k_\gamma\,\slashed{\epsilon}^{\gamma *}_\perp\frac{\slashed{\bar{n}}}{2}u_\xi^{(s)}\,.
\end{equation}
According to~\eqref{THH}, only the restricted cuts have to be considered. For the photon cut, the expression is needed only to $\mathcal{O}(\alpha_{\rm em})$, whereas the cut strange-quark propagator, which corresponds to the jet function, must be kept in a form valid to all orders in $\alpha_s$ (see~\eqref{jetmomentum}). It is sufficient to consider the Dirac structure of the combined operator matrix element. This leads to the following definition of the anti-hard-collinear jet function in terms of the two matching coefficients, up to a prefactor $\kappa$:
\begin{equation}
\label{antijet2}
J^{(\bar{n})}(u,\omega_1)= \kappa
[(D-2)J^{(\bar{n}),a}+(4-D)J^{(\bar{n}),b}]\,, 
\end{equation} 
where $D=4-2\eps$. The second contribution turns out to be evanescent.

We can cross-check this general result at leading order in two steps. We first compute the anti-hard-collinear jet function from its definition in the factorisation formula,~\eqref{antijet1}. We then compute it from the matching condition~\eqref{antijetmatch}, which fixes $J^{(\bar{n}),a}$ and $J^{(\bar{n}),b}$. Comparing the two fixes the prefactor to $\kappa =1$. We find full agreement between the two definitions of the anti-hard-collinear jet function at leading order (see Appendix~\ref{consistencycheck} for details).

One subtlety remains. The definition~\eqref{antijet2}, based on the matching condition~\eqref{antijetmatch}, uses the gauge-invariant soft gluon building block $\mathcal{A}^\mu_{s}=S^\dagger_{\bar n}[iD^\mu_s S_{\bar n}]$ defined in~\eqref{eq:soft_building_block}. The factorisation formula, by contrast, uses the field-strength tensor $G_s^{\mu_\perp\nu}$, consistent with the definition of the shape function $g_{17}$ in~\eqref{shapedefinition}. The two are related by the identity\footnote{Note that the indices of $G_s$ are swapped between the left- and right-hand sides; the minus sign follows from the antisymmetry of the field-strength tensor.}
\begin{equation}
\label{eq:A_vs_G}
\mathcal{A}^\mu_{\perp,s}(x_+)=\int_{-\infty}^0 ds\,\big(S^\dagger_{\bar n}\,g_s\bar n_\nu G_s^{\nu\mu_\perp}S_{\bar n}\big)(x_++s\bar n)
=-\frac{i}{i\bar{n}\cdot \partial+i0}\big(S^\dagger_{\bar{n}}\,g_s\bar{n}_\nu G_s^{\mu_\perp\nu}S_{\bar n}\big)(x_+)\,,
\end{equation}
which follows from $\bar n\cdot D_s S_{\bar n}=0$ and $[iD_s^\nu,iD_s^\mu]=ig_sG_s^{\nu\mu}$, implying $\bar n\cdot\partial\,\mathcal{A}^\mu_{s}=S^\dagger_{\bar n}\,g_s\bar n_\nu G_s^{\nu\mu}S_{\bar n}$. Hence the matching condition expressed in terms of $G_s^{\mu_\perp\nu}$ yields an additional factor $\frac{1}{\omega_1+i0}$, corresponding to $\frac{1}{i\bar{n}\cdot \partial+i0}$. This factor has to be included in the definition~\eqref{antijet2}.

In summary, we have established a general definition of the anti-hard-collinear jet function, valid to all orders in $\alpha_s$, which allows it to be extracted uniquely from the $b \to s \gamma g$ amplitude induced by the operator $Q_1^q$.

\section{NLO corrections known in the literature}
\label{sec:literature}
Among the functions entering the factorisation formula~\eqref{FactorisationFormula}, two have already been computed beyond leading order in previous studies. For the quark jet function $J^{(n)}$, the NLO and NNLO results are given in~\cite{Becher:2006qw}, and the three-loop (N$^3$LO) result was obtained in~\cite{Bruser:2018rad}. For the soft function $g_{17}$, which is non-perturbative, we determined the one-loop anomalous dimension in~\cite{Bartocci:2024bbf}. There we also solved the associated renormalisation-group equation analytically in closed form. In the following we collect only those results that are needed for the pole-cancellation analysis of this paper.

\subsection{Quark jet function}
\label{subsec:quarkjetfunction}
The hard-collinear jet function is obtained from the two-point correlator of gauge-invariant hard-collinear quark fields, which in momentum space takes the form
\begin{equation}\label{eq:Jscet}
   \delta^{ij}\,\frac{\slashed{n}}{2}\,\bar{n}\cdot p\,{\cal J}^{(n)}(p^2;\mu) = \int d^4x\,e^{-ip\cdot x}\,\langle 0|T\{\chi^i_{hc}(0)\,\overline\chi^j_{hc}(x)\}|0\rangle \,,
\end{equation} 
with $\chi_{hc}$ the gauge-invariant field of the massless strange quark. Taking the imaginary part, or equivalently the discontinuity, of this correlator yields the quark jet function $J^{(n)}$,
\begin{equation}
   J^{(n)}(p^2;\mu) = \frac{1}{\pi}\,{\rm Im}\left[ i{\cal J}^{(n)}(p^2;\mu) \right]
   =  \frac{1}{2 i \pi}\, {\rm Disc}\left[ i{\cal J}^{(n)}(p^2;\mu)    \right] = \delta(p^2) + {\cal O}(\alpha_s) \,.
\end{equation}
The unrenormalised one-loop correction was computed in~\cite{Bosch:2004th} and reads
\begin{align}
\label{eq:Jbare}
   J^{(n),(1)}_{\rm bare}(p^2;\mu)
   = \, &\frac{\alpha_s C_F}{4\pi} \Bigg[
   \left( \frac{4}{\eps^2} + \frac{3}{\eps} + 7 - \pi^2 \right)
   \delta(p^2)
   + 4 \left( \frac{\ln{p^2/\mu^2}}{p^2} \right)_{\!*}^{\![\mu^2]}
   - \left( \frac{4}{\eps} + 3 \right)
   \left( \frac{1}{p^2} \right)_{\!*}^{\![\mu^2]} \Bigg] \,,
\end{align}
where the star-distributions are those of Appendix~\ref{app:distribution_definition}.
In the $\overline{\text{MS}}$ scheme, renormalisation proceeds through the convolution
\begin{equation}
    J^{(n)}(p^2;\mu) = \int dp^{\prime \, 2} \, Z_J(p^2,p^{\prime \, 2};\mu) \, J^{(n)}_{\rm bare}(p^{\prime \, 2}) \,, 
\end{equation}
in which the bare result~\eqref{eq:Jbare} is combined with the renormalisation factor at one loop, known from~\cite{Becher:2006qw},
\begin{equation}
\label{eq:Z-jet}
   Z_J(p^2,p^{\prime \, 2};\mu) = \delta(p^2 - p^{\prime \, 2}) + \frac{\alpha_s C_F}{4\pi} \left[ \delta(p^2 - p^{\prime \, 2}) \left( - \frac{4}{\eps^2} - \frac{3}{\eps} \right) + \frac{4}{\eps} \left( \frac{1}{p^2 - p^{\prime \, 2}} \right)_{\!*}^{\![\mu^2]} \right] + \, \mathcal{O}(\alpha_s^2) \,.
\end{equation}

\subsection{Subleading shape function \texorpdfstring{$g_{17}$}{g17}}
\label{subsec:g17}
The soft dynamics of the resolved contribution is encoded in the subleading shape function $g_{17}$. It is given by the double Fourier transform of a forward matrix element taken between static $\bar B$-meson states,
\begin{align}\label{eq:g17def}
   g_{17}(\omega,\omega_1;\mu) = \frac{1}{2M_B} \int\frac{dr}{2\pi}\,e^{-i\omega_1 r}
    \int\frac{dt}{2\pi}\,e^{-i\omega t} \, \langle\bar B_v| \mathcal{O}_{17}(t,r) |\bar B_v\rangle \,,
\end{align}
where the relevant non-local HQET operator is~\cite{Benzke:2010js,Bartocci:2024bbf}
\begin{equation}
\label{eq:Q17}
    \mathcal{O}_{17}(t,r) = \big(\bar h_v S_n\big)_-(tn)\,
    \nbslash(1+\gamma_5) \big(S_n^\dagger S_{\bar n}\big)_+(0)\,
    i\gamma_\alpha^\perp\bar n_\beta\,
    \big(S_{\bar n}^\dagger\,g_s G_s^{\alpha\beta} S_{\bar n} 
    \big)_+(r\bar n)\,
    \big(S_{\bar n}^\dagger h_v\big)_+(0) \,.
\end{equation}
The Wilson lines along $n$ originate from decoupling soft gluons from the hard-collinear strange-quark propagator, while those along $\bar n$ arise in the same way from the anti-hard-collinear propagators of the quark loop. The subscripts ``$+$'' and ``$-$'' refer to the Keldysh (closed-time-path) formalism~\cite{Schwinger:1960qe,Keldysh:1964ud}, summarised for instance in~\cite{Becher:2007ty}. Fields labelled ``$+$'' are part of the amplitude and are treated with the usual HQET Feynman rules for time-ordered products. Fields labelled ``$-$'' are part of the conjugate amplitude and are treated with the complex-conjugate rules for anti-time-ordered products.

We determined the one-loop renormalisation factor $Z_{17}$ of this operator, which extends along two light-cone directions, in~\cite{Bartocci:2024bbf}. It naturally splits into a colour structure proportional to $C_F$ and one proportional to $C_A$,
\begin{equation}
\label{eq:Z-g17}
    Z_{17}(\omega,\omega_1,\omega',\omega'_1;\mu) = \delta(\omega - \omega') \delta(\omega_1 - \omega'_1) + Z_{17}^{(1)}\Big\vert_{C_F} + Z_{17}^{(1)}\Big\vert_{C_A} + \mathcal{O}(\alpha_s^2) \,.
\end{equation}
The $C_F$ term is identical to the renormalisation factor of the leading-power shape function~\cite{Bosch:2004th},
\begin{align}
\label{eq:CFZfactor}
    Z_{17}^{(1)}\Big\vert_{C_F}  = 
     \frac{\alpha_s C_F}{4\pi} \delta(\omega_1 - \omega'_1) \bigg\{ \left(\frac{2}{\eps^2} + \frac{4}{\eps} \ln \frac{\mu}{\bar{\Lambda}-\omega} - \frac{2}{\eps} \right) \delta(\omega-\omega')
    - \frac{4}{\eps} \left[ 
\frac{\theta(\omega'-\omega)}{\omega'-\omega} \right]_+^{(\bar{\Lambda})} \bigg\} \,, 
\end{align}
with $\bar\Lambda = M_B - m_b$ the mass difference between the $B$ meson and the $b$-quark pole mass. The $C_A$ term, which has no counterpart at leading power, is
\begin{align}
\label{eq:CAZfactor}
    Z_{17}^{(1)}\Big\vert_{C_A} = \, &\frac{\alpha_s C_A}{4\pi} \delta(\omega - \omega') 
    \bigg\{ \left( \frac{1}{\eps^2} + \frac{1}{\eps}\ln \frac{\mu^2}{\omega_1^2}\right) \delta(\omega_1-\omega'_1) \nonumber \\[0.2em]
    + \, &\frac{2}{\eps} \frac{\omega_1}{(\omega'_1)^2} \big[ \theta(\omega_1)\theta(\omega'_1 - \omega_1) - \theta(-\omega_1) \theta(\omega_1 - \omega'_1) \big] - \frac{1}{\eps} \, \text{Re} \, \mathbf{H}(\omega_1,\omega'_1) \bigg\} \,.
\end{align}
All distributions appearing above, including $\mathbf{H}$, are defined in Appendix~\ref{app:distribution_definition}.

\section{Hard functions at NLO}
\label{sec:hardmatching}
\begin{figure}[t]
    \centering
     \includegraphics[width=0.3\textwidth]{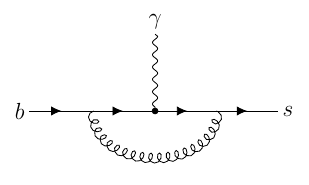}
    \caption{The one-loop hard-matching correction contributing to $H_{7\gamma}$. Field renormalisation diagrams are not shown.}
    \label{fig:hard_7}
\end{figure}

The hard matching functions entering {the} factorisation formula are obtained by matching the electroweak operators onto SCET. We determine their IR poles, which are the relevant input for the pole-cancellation test in Section~\ref{sec:pole_cancellation}. They also  determine the RG  equations which allow one to resum large logarithms between the hard and hard-collinear scales (see~\cite{Neubert:2004dd,Becher:2006pu}).

The hard function associated with $Q_{7\gamma}$ at NLO  can be extracted from the $\mathcal{O}(\alpha_s)$ corrections to the on-shell partonic matrix element for the process $b \to s \gamma$. For on-shell external states, all loop corrections to the SCET matrix element are scaleless and vanish. Hence, only the 
$\alpha_s$ corrections to the QCD matrix element have to be calculated. The result for the diagonal contribution of the operator $Q_{7\gamma}$ reads (see Figure~\ref{fig:hard_7}) 

\begin{align}
\label{eq:H7poles}
\langle Q_{7\gamma}\rangle^{(1)}_{\text{bare}}
= -\frac{\alpha_s C_F}{4\pi}
\Bigg(
&\frac{1}{\eps_{\IR}^2}
+ \frac{5}{2\eps_{\IR}}
+ \frac{1}{\eps_{\IR}} \ln\frac{\mu^2}{m_b^2}
+ \frac{1}{\eps_{\UV}}  \nonumber \\
+ &\frac{1}{2}\ln^2\frac{\mu^2}{m_b^2}
+ \frac{5}{2}\ln\frac{\mu^2}{m_b^2}
+ \ln\frac{\nu^2}{m_b^2}
+ 6
+ \frac{\pi^2}{12}
\Bigg)
\langle \mathcal{O}_{7\gamma}\rangle^{(0)} \,.
\end{align}
We note that the field renormalisation of the external lines is already included here.  The superscript $(1)$ denotes the first order in $\alpha_s$ and $\nu$ refers to the scale of the Wilson coefficients which we distinguish from the
factorisation scale $\mu$. Correspondingly, we have also distinguished the IR poles $\eps_{\IR}$ from the UV poles $\eps_{\UV}$.

We explicitly extract  the corresponding IR divergences as a $Z$ factor of the hard function, upon standard subtraction of the UV singularities\footnote{We note that  $Z_{7\gamma}(\mu)$ is  the ultraviolet $Z$ factor of the effective operator  ${\cal O}_{7\gamma}$~\cite{toappearRB}.}
\begin{equation}
\label{ZQ7IR} 
    Z_{7\gamma}^{-1}(\mu)= 1 + \frac{\alpha_s C_F}{4\pi}\left(\frac{1}{\eps^2}+\frac{5}{2\eps}+\frac{1}{\eps}\ln{\frac{\mu^2}{m_b^2}}\right)\,.
\end{equation}

From this we can now read off the diagonal part of the UV- and IR-finite hard matching coefficient. Quoting also the IR-finite non-diagonal mixing contributions from the operators $Q_{8g}$ and $Q_1^q$ from~\cite{Neubert:2004dd}, the result reads
\begin{eqnarray}
\label{Hgamma}
   H_{7\gamma}(\mu,\nu)
   &=& C_{7\gamma}(\nu) \left[ 1
    - \frac{\alpha_s(\nu) C_F}{4\pi} \left( \frac12 \ln^2\frac{\mu^2}{m_b^2}
    + \frac52 \ln\frac{\mu^2}{m_b^2}+ \ln\frac{\nu^2}{m_b^2} + 6 + \frac{\pi^2}{12} \right) \right] \nonumber\\
   &&\mbox{}+ C_{8g}(\nu)\,\frac{\alpha_s(\nu) C_F}{4\pi}
    \left( \frac43 \ln\frac{\nu^2}{m_b^2} + \frac{11}{3} - \frac{2\pi^2}{9}
    + \frac{2\pi i}{3} \right) \nonumber\\
   &&\mbox{}+ C_1^q(\nu)\,\frac{\alpha_s(\nu) C_F}{4\pi}
    \left( - \frac{52}{27} \ln\frac{\nu^2}{m_b^2} + \tilde{g}(z)
    + \varepsilon_{\rm CKM}\,\big[ \tilde{g}(0) - \tilde{g}(z) \big] \right) \,,
\end{eqnarray} where $\varepsilon_{\rm CKM} = - (V_{ub} V_{us}^*)/(V_{tb} V_{ts}^*)$ and $z=(m_c/m_b)^2$ is the argument of the penguin function $\tilde{g}(z)$~\cite{Greub:1996tg}. Since the Wilson coefficients do not depend on the quark flavour (see Appendix~\ref{app:setup}), the contributions with $q=c$ and $q=u$ have been combined by means of $\varepsilon_{\rm CKM}$. The penguin function reads
\begin{eqnarray}\label{gz}
   \tilde{g}(z) &=& - \frac{833}{162} - \frac{20\pi i}{27} + \frac{8\pi^2}{9}\,z^{3/2}
    \nonumber\\
   &&\mbox{}+ \frac{2z}{9} \left[ 48 - 5\pi^2\! - 36\zeta_3
    + (30\pi - 2\pi^3) i + \!(36 - 9\pi^2 + 6\pi i) \ln z
    + \!(3 + 6\pi i) \ln^2 z + \ln^3 z \right] \nonumber\\
   &&\mbox{}+ \frac{2z^2}{9} \left[ 18 + 2\pi^2 - 2\pi^3 i
    + (12 - 6\pi^2) \ln z + 6\pi i\ln^2 z + \ln^3 z \right] \nonumber\\
   &&\mbox{}+ \frac{z^3}{27} \left[ -9 - 14\pi^2 + 112\pi i
    + (182 - 48\pi i) \ln z - 126\ln^2 z \right] + \mathcal{O}(z^4) \,. 
\end{eqnarray}
We emphasise that in our analysis we adopt the scaling $m_c^2 = \mathcal{O}(m_b \Lambda_{\rm QCD})$ and therefore set $z=0$ in $H_{7\gamma}(\mu,\nu)$ for consistency.\footnote{However, one could add terms of order $z^1=m_c^2/m_b^2$ or $z^2$ as power corrections. Such terms are numerically important and will be included in the phenomenological analysis; in the massless case considered here, $z=0$.}

Similarly, we compute the hard matching for the two electroweak operators $Q_{1}^q$ and  $Q_{2}^q$. As in Section~\ref{sec:antijet_from_matching}, we use the basis~\eqref{eq:Q1-Q2}, matched onto the SCET operators~\eqref{eq:O1-matching}.

This basis has the advantage that the SCET operator ${\cal O}_2^q$ does not contribute to the anti-hard-collinear jet function due to its colour structure. Indeed, this jet function is generated by a quark loop emitting a photon and a gluon, followed by a gluonic loop. In ${\cal O}_2^q$ the two anti-hard-collinear quark fields form a colour singlet, so that the quark loop closes onto a colour trace; the emission of the soft gluon then yields ${\rm Tr}(t^a)=0$, and its contribution vanishes.

\begin{figure}[t!]
    \centering
     \includegraphics[width=0.6\textwidth]{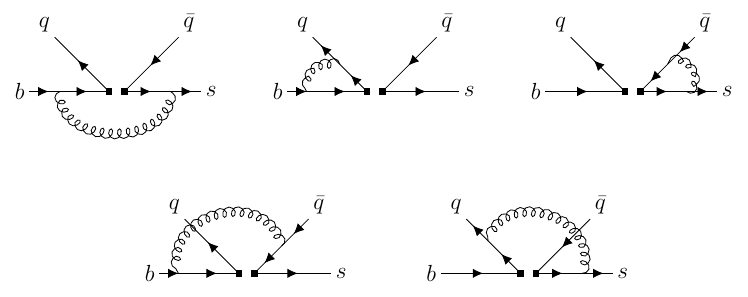}
    \caption{One-loop diagrams contributing to the hard matching function for the operators $Q_{1}^q$ and $Q_{2}^q$. Contributions due to field         renormalisation are not shown.}
    \label{fig:hard_1}
\end{figure}

The hard-matching coefficients can be extracted from the QCD $b \to s \bar{q} q$ matrix element with on-shell partons, where the outgoing quark (anti-quark) carries the momentum fraction $u$ ($\bar{u} = 1-u$) of the photon momentum, owing to the delocalisation of the two anti-hard-collinear quark fields on the light cone within the SCET operators.  At order $\alpha_s$ the diagrams in Figure~\ref{fig:hard_1} yield
\begin{align}
\label{eq:H1_bare_full}
\langle Q_1^{q}\rangle^{(1)}_{\text{bare}} &= \frac{\alpha_s}{4\pi} \Big[ \big( C_F f_1(u) + C_A f_2(u) \big) \langle \mathcal{O}_1^{q}\rangle^{(0)} - f_2(u)\, \langle \mathcal{O}_2^{q}\rangle^{(0)} \Big] \,, \\[0.3em]
\langle Q_2^{q}\rangle^{(1)}_{\text{bare}} &= \frac{\alpha_s}{4\pi} \Big[ - f_3(u)\, \langle \mathcal{O}_1^{q}\rangle^{(0)} + \big( C_F f_1(u) + C_A f_3(u) \big) \langle \mathcal{O}_2^{q}\rangle^{(0)} \Big] \,. \nonumber
\end{align}
{Also here the field renormalisation of the external fields is already included.} The superscript $(1)$ denotes the first order in $\alpha_s$. The loop functions depend on the light-cone momentum fraction $u$ carried by the anti-hard-collinear quark in the SCET operator and read
\begin{align}
f_1(u) = &
-\frac{1}{\eps^2_{\IR}}
-\frac{1}{\eps_{\IR}}
\ln\frac{\mu^2}{m_b^2}
-\frac{5}{2\eps_{\IR}}-\frac{6}{\eps_{\UV}}
-\frac12 \ln^2 \frac{\mu^2}{m_b^2}
-\frac{5}{2}\ln \frac{\mu^2}{m_b^2} -6 \ln \frac{\nu^2}{m_b^2} \nonumber
+g(u) -\frac{\pi^2}{12} -16 \,,\\ 
\begin{split}
f_2(u)=&-\frac{1}{\eps_{\IR}^2}
- \frac{1}{\eps_{\IR}}
 \ln\!
\frac{\mu^2}{m_b^2 u\bar{u}}- \frac{i\pi}{\eps_{\IR}}
-\frac{3}{2\eps_{\IR}}+\frac{3}{\eps_{\UV}}
-\frac{1}{2} \ln^2\!
\frac{\mu^2}{m_b^2 u \bar{u}}
-\frac{3}{2} \ln \frac{\mu^2}{m_b^2}
+3 \ln \frac{\nu^2}{m_b^2} -i\pi\ln\frac{\mu^2}{m_b^2u\bar{u}}
\\[6pt]
&+\frac32 \ln^2\bar{u}+\frac{7\pi^2}{12}
+3-f(\bar{u})
-3 (1+i\pi)\ln\bar{u}\,,\nonumber
\end{split}\\
f_3(u) = &\frac{3}{\eps_{\UV}}
+3\ln\!\frac{\nu^2}{m_b^2}+5-\frac{1}{2}g(u) \,. 
\end{align}
The two functions $f(u)$ and $g(u)$ are defined as~\cite{Beneke:2020vnb},
 \begin{align}
     f(u) = & {\rm Li}_2(1-u) + 2 \ln^2 u - (3+2i\pi) \ln u -\frac{\ln u }{1-u}\,,
\nonumber\\
g(u) = & 3 \left(\frac{1-2u}{1-u} \ln u - i \pi \right) + \bigg[2 \,{\rm Li}_2(u) - \ln^2 u + \frac{2 \ln{u}}{1-u} - (3+2 i\pi)\ln{u} - \,(u \to  1-u ) \bigg]\,.
\label{eq:fdef}
\end{align}
Furthermore, we set $E_\gamma = m_b/2$ in the above expressions. As before, we separated UV singularities and the associated scale $\nu$ of the Wilson coefficients $C_{1,2}^q(\nu)$ from IR singularities and the factorisation scale $\mu$.  

The hard function $H^q_1$ entering the factorisation formula is given at order $\alpha_s$ by the sum of all on-shell amplitudes in~\eqref{eq:H1_bare_full} proportional to the tree-level matrix element of ${\cal O}^q_1$, because the on-shell $b \to s \gamma g$ matrix element of the operator ${\cal O}^q_2$ vanishes. It consists of a diagonal part proportional to $C_1^q$ and a non-diagonal part proportional to $C_2^q$,
\begin{equation}
    \label{eq:H1_bare_function}H^q_1(u,\mu,\nu)_{\rm bare}   = C_1^q(\nu) \Big[ 1 +  \frac{\alpha_s}{4 \pi} \Big( C_F f_1(u) + C_A f_2(u)\Big) \Big] + C_2^q(\nu) \Big[ -\frac{\alpha_s}{4 \pi} f_3(u) \Big] + \mathcal{O}(\alpha_s^2)\,.
\end{equation}
Again, the field renormalisation of the external lines is already included here.  We  explicitly extract  the corresponding IR divergences as a $Z$ factor of the diagonal part of the hard function; the non-diagonal part is IR finite:\footnote{$Z_{11}(u;\mu)$ corresponds to the diagonal entry of the UV $Z$ matrix of the effective operators ${\cal  O}_1^q$ and ${\cal O}_2^q$ (see for more details~\cite{toappearRB}).}
\begin{equation}
\label{eq:Z-factor_hard_Q1} 
    Z_{11}^{-1}(u;\mu)=  1
  + \frac{\alpha_s C_F}{4\pi} \left( \frac{1}{\eps^2} + \frac{1}{\eps} \ln \frac{\mu^2}{m_b^2} + \frac{5}{2\eps} \right)   
+ \frac{\alpha_s C_A}{4\pi} \left( \frac{1}{\eps^2} + \frac{1}{\eps} \ln \frac{\mu^2}{m_b^2 u \bar{u}} + \frac{i \pi}{\eps} + \frac{3}{2\eps} \right)\,.
\end{equation}

After UV renormalisation, one obtains the finite hard functions. In this step, both the operator renormalisation and the renormalisation of the external fields give rise to additional finite contributions from evanescent operators (see~\cite{toappearRB} for details).

Since the factorisation formula~\eqref{FactorisationFormula} involves the product $H_1^q(u)\,H_{7\gamma}^*$, the IR poles of the complete hard function are collected in the combined $Z$ factor
\begin{align}
\label{eq:Z_h}
    Z_h(u;\mu) \equiv Z_{11}^{-1}(u;\mu)\,\big[Z_{7\gamma}^{-1}(\mu)\big]^* = 1
    &+ \frac{\alpha_s C_F}{4\pi} \left( \frac{2}{\eps^2} + \frac{2}{\eps} \ln \frac{\mu^2}{m_b^2} + \frac{5}{\eps} \right) \nonumber \\
    &+ \frac{\alpha_s C_A}{4\pi} \left( \frac{1}{\eps^2} + \frac{1}{\eps} \ln \frac{\mu^2}{m_b^2 u \bar{u}} + \frac{i \pi}{\eps} + \frac{3}{2\eps} \right) + \mathcal{O}(\alpha_s^2)\,,
\end{align}
which is the input for the pole-cancellation test in Section~\ref{sec:pole_cancellation}.

\section{Anti-hard-collinear jet function at NLO}
\label{sec:antijet_function}

The anti-hard-collinear jet function can be extracted from the anti-hard-collinear momentum region of the $b\to s\gamma g$ amplitude:
\begin{equation}
b(m_b v) \to s(E_X n)\gamma(E_\gamma \bar{n}) g\left(\omega_1 n/2 \right),
\end{equation}
where $E_X$ denotes the energy of the hard-collinear strange quark,
induced by the operator $Q_1^q$. This extraction is uniquely fixed by the matching calculation presented in Section~\ref{sec:antijet_from_matching}.

To this end, we compute the $\mathcal{O}(\alpha_s)$ corrections to the amplitude and isolate the contribution arising from the anti-hard-collinear region. We present the results for a massless up quark ($q=u$) in the loop at both LO and NLO.
A complete phenomenological analysis requires the inclusion of charm-quark mass effects, which we defer to future work. However, the $1/\eps$ divergences are fixed by the anomalous dimensions of the functions entering the factorisation formula, which are independent of the charm-quark mass. The massless approximation therefore correctly reproduces the full structure of $1/\eps$ divergences. It is therefore sufficient for establishing the cancellation of poles among the four functions entering the factorisation formula, providing a non-trivial consistency check of both the factorisation framework and our calculations.

\subsection{The \texorpdfstring{$b\to s \gamma g$}{b -> s gamma g} amplitude at one loop}
\begin{figure}[t]
    \centering
    \includegraphics[width=0.55\textwidth]{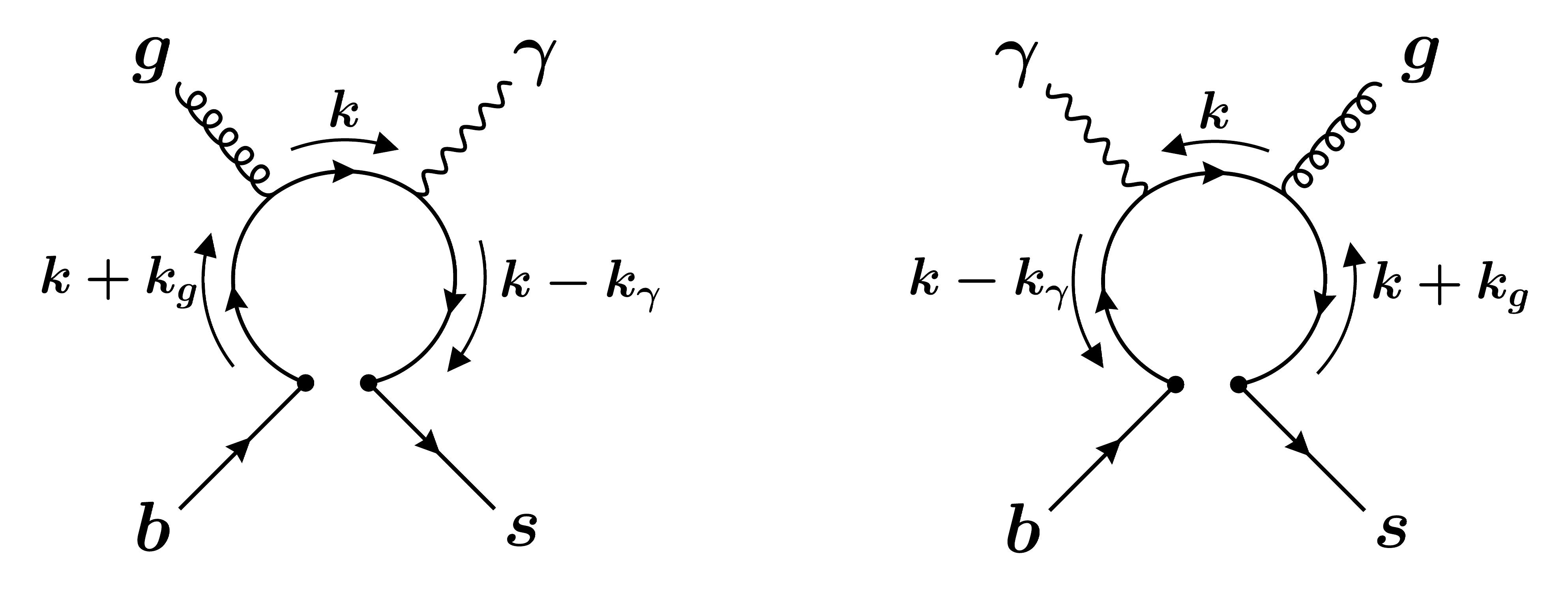}
    \caption{LO diagrams contributing to the $b \to s\gamma g$ amplitude. The two diagrams differ by the interchange of the gluon and photon emission vertices.}
    \label{fig:antijet_LO_diagrams}
\end{figure}

We begin by presenting the LO {amplitude}, which allows us to establish our notation and computational setup, and to highlight the main structural features that persist in the more involved NLO calculation.
As shown in Figure~\ref{fig:antijet_LO_diagrams}, the $b\to s \gamma g$ amplitude induced by the operator $Q_1^q$ arises only at the loop level. Consequently, the LO contribution corresponds to a one-loop calculation, while the NLO result requires the evaluation of two-loop diagrams. In this calculation, the gluon and photon momenta are given by $k_g^\mu=\omega_1\frac{n^\mu}{2}$ and $k_\gamma^\mu=E_\gamma \bar{n}^\mu $, while the loop momentum can be decomposed in the standard form as $k^\mu=k_+\frac{\bar{n}^\mu}{2}+k_-\frac{n^\mu}{2}+k_\perp^\mu$. 

\begin{figure}[t]
    \centering
    \includegraphics[width=0.72\textwidth]{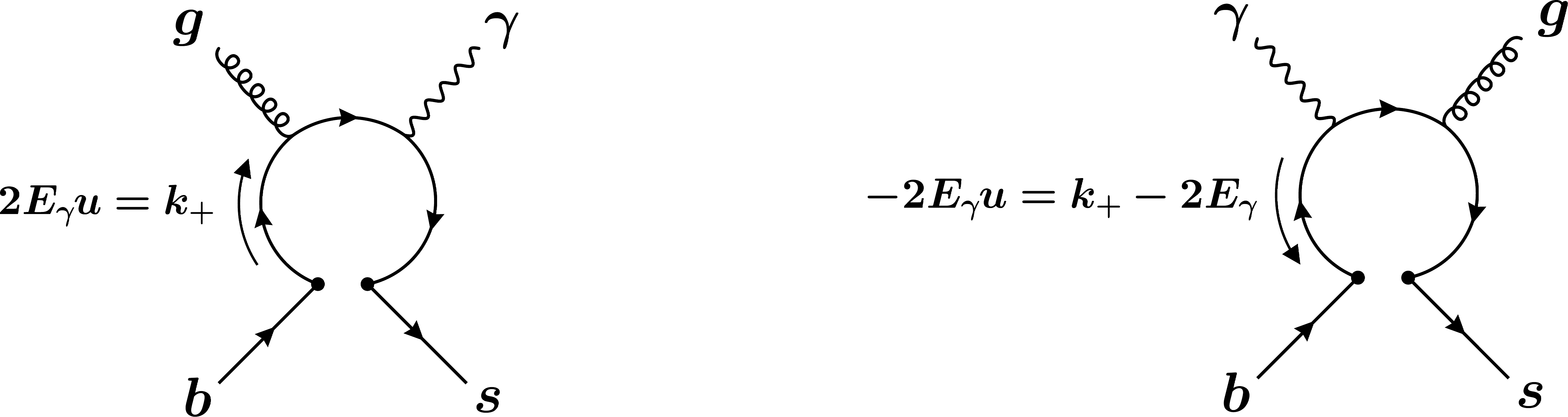}
    \caption{Schematic representation of the anti-hard-collinear momentum fraction $u$. The momentum assignments in both diagrams follow those of Figure~\ref{fig:antijet_LO_diagrams}.} 
    \label{fig:u-dependence_LO}
\end{figure}

Additionally, we have to determine the explicit dependence on the momentum fraction $u$ induced by the delocalisation of anti-hard-collinear quark fields from hard gluon exchanges. The variable $u$ represents the anti-hard-collinear momentum fraction flowing out of the four-quark vertex, as illustrated in Figure~\ref{fig:u-dependence_LO}.  

In particular, in the left diagram we identify $u = \frac{k_+}{2E_\gamma}$, while in the right diagram we use the complementary fraction $\bar{u} \equiv 1 - u $. We impose these kinematic relations by introducing a loop-momentum-dependent $\delta$ function in the amplitudes.
The one-loop amplitudes related to the diagrams in Figure~\ref{fig:antijet_LO_diagrams} read
\begin{align}
\label{eq:one_loop_u_massless_1}
    \begin{split}
       \mathcal{A}_{\text{LO}}^{(1),u}(u) = \tilde{\mu}^{2\eps} \int & \frac{d^D k}{(2\pi)^D} \, \bar{u}^{(s)}_{hc,L} \gamma^\mu i \frac{\slashed{k} - \slashed{k}_\gamma}{(k - k_\gamma)^2 + i0} i Q_u \slashed{\epsilon}_\perp^*(k_\gamma) i \frac{\slashed{k}}{k^2 + i0} \\
        & \times i g_s \slashed{\epsilon}_{s,\perp}^*(k_g) i \frac{\slashed{k} + \slashed{k}_g}{(k + k_g)^2 + i0} \gamma_\mu b_L \, \delta \left( u - \frac{k_+}{2 E_\gamma} \right),
    \end{split}
\end{align}

\begin{align}
\label{eq:one_loop_u_massless_2}
    \begin{split}
    \mathcal{A}_{\text{LO}}^{(2),u}(u) = - \tilde{\mu}^{2\eps} \int & \frac{d^D k}{(2\pi)^D} \, \bar{u}^{(s)}_{hc,L} \gamma^\mu i \frac{\slashed{k} + \slashed{k}_g}{(k + k_g)^2 + i0} i g_s \slashed{\epsilon}_{s,\perp}^*(k_g) i \frac{\slashed{k}}{k^2 + i0} \\
        & \times i Q_u \slashed{\epsilon}_\perp^*(k_\gamma) i \frac{\slashed{k} - \slashed{k}_\gamma}{(k - k_\gamma)^2 + i0} \gamma_\mu b_L \, \delta \left( \bar{u} - \frac{k_+}{2 E_\gamma} \right).
    \end{split}
\end{align}
Here, $\Bar{u}^{(s)}_{hc,L}$ and $b_L$ are the spinors for the left-handed hard-collinear strange quark and the bottom quark, respectively, and $\tilde\mu^2=\mu^2 e^{\gamma_E}/(4\pi)$ denotes the $\overline{\rm MS}$ scale. The polarisation vectors $\epsilon^*_{\perp}(k_\gamma)$ and $\epsilon_{s,\perp}^*(k_g)=t^a\epsilon_{s,\perp}^{a*}(k_g)$ correspond to the photon and the soft gluon, respectively. Since both the photon and the gluon are on-shell, we choose their polarisation vectors to be transverse to $n^\mu$ and $\bar{n}^\mu$.

To employ the standard reduction techniques, we rewrite the delta function using the Sokhotski–Plemelj formula:
\[
\frac{1}{x \pm i0} = \mathcal{P}\left(\frac{1}{x}\right) \mp i \pi \delta(x),
\]
which allows us to express the delta function as a difference of linear propagators:
\begin{equation}
\label{eq:delta_linear}
    \delta\left(u-\frac{k_+}{2E_\gamma}\right)=\frac{2E_\gamma}{2\pi i}\left(\frac{1}{k_+-2E_\gamma u-i0}-\frac{1}{k_+-2E_\gamma u+i0}\right)\,.
\end{equation}

The diagrams in Figure~\ref{fig:antijet_LO_diagrams} receive contributions from only one region, namely the anti-hard-collinear region. Therefore, the loop integrals can be computed directly, without performing any expansion. Summing the two amplitudes, we obtain:
\begin{align}
\label{eq:one_loop_massless_udep}
    \begin{split}
          \mathcal{A}_{\text{LO}}^{u}(u)=  -&\frac{g_sQ_uE_\gamma}{
          4\pi^{2}}\left(\frac{-2E_\gamma \omega_1}{\mu^2}\right)^{-\eps}e^{\gamma_E\eps}\frac{\Gamma(1+\eps)}{1-\eps}[u(1-u)]^{-\eps}\theta(u,1-u)\\
            \times&\Bar{u}^{(s)}_{hc,L}\left[\slashed{\epsilon}_{s,\perp}^*(k_g)\slashed{\epsilon}_{\perp}^*(k_\gamma)(1-u(1-\eps))- \slashed{\epsilon}_{\perp}^*(k_\gamma)\slashed{\epsilon}_{s,\perp}^*(k_g)(u+(1-u)\eps)\right]\frac{\slashed{\bar n}}{2}b_L\,,
             \end{split}
                \end{align}
where $\theta(u,1-u)\equiv\theta(u)\,\theta(1-u)$. We observe that the $u$-dependence of the two Dirac structures differs, and that the amplitude is anti-symmetric under the simultaneous replacement $u \to \bar{u}$ and interchange of the two Dirac structures. 
Furthermore, the dependence on $E_\gamma$ and $\omega_1$ factorises completely and can be deduced from dimensional analysis of the amplitude. 
In particular, the expression includes the combination $\left(\frac{-2E_\gamma(\omega_1+i0)}{\mu^2}\right)^{-\eps}$, where the minus sign, also expected at NLO, ensures the correct imaginary part, which will prove crucial for the cancellation of poles in this paper. Throughout the following part of this section, the $i0$ prescription is understood. Lastly, we note that the amplitude at one loop is finite in the limit $\eps\to 0$. 
Integrating the above expression over the variable $u$ reproduces the known result for the standard amplitude without $u$-dependence:
\begin{align}
\label{eq:one_loop_massless_result_finite}
    \begin{split}
    \mathcal{A}_{\text{LO}}(b\to s\gamma g) = \int_0^1du\,\mathcal{A}^u_{\text{LO}}(u)=&-\frac{g_sQ_u E_\gamma}{8\pi^2}\left(\frac{-2E_\gamma\omega_1}{\mu^2}\right)^{-\eps}e^{\gamma_E\eps}(1+\eps)\frac{\Gamma(1+\eps)\Gamma(1-\eps)^2}{(1-\eps)\Gamma(2-2\eps)}\\& \times \Bar{u}^{(s)}_{hc,L} \left[\slashed{\epsilon}_{s,\perp}^*(k_g),\slashed{\epsilon}_{\perp}^*(k_\gamma)\right]\frac{\slashed{\bar{n}}}{2} b_L \,. 
    \end{split}
\end{align}
Now we can use the matching calculation in~\eqref{antijetmatch} to project the anti-hard-collinear function at LO out of the $b\to s\gamma g$ amplitude. 
 We find for the $u$-dependent case:
\begin{align}
\label{eq:one_loop_udep_antijet}
    \begin{split}
          J_{\text{LO}}^{(\bar{n})}(u,\omega_1;\,\mu)=  -&\frac{E_\gamma}{
          4\pi^{2}}\left(\frac{-2E_\gamma \omega_1}{\mu^2}\right)^{-\eps}\frac{1}{\omega_1+i0}e^{\gamma_E\eps}\frac{\Gamma(1+\eps)}{1-\eps}[u(1-u)]^{-\eps}\theta(u,1-u)\\
            \times&\left[2\eps(1-u(1-\eps))- (2-2\eps)(u+(1-u)\eps)\right]\,,
             \end{split}
                \end{align}
where, as explained in Section~\ref{sec:antijet_from_matching}, an additional factor $\frac{1}{\omega_1+i0}$ has been added.
Integrating now in $u$ we get the $u$-independent anti-hard-collinear jet function:
\begin{align}
    \label{eq:antijet_LO_u_integrated}
    \begin{split}
    J_{\text{LO}}^{(\bar{n})}(\omega_1;\,\mu) = &-\frac{E_\gamma}{8\pi^2}\left(\frac{-2E_\gamma\omega_1}{\mu^2}\right)^{-\eps}\frac{1}{\omega_1+i0}e^{\gamma_E\eps}(1+\eps)\frac{\Gamma(1+\eps)\Gamma(1-\eps)^2}{(1-\eps)\Gamma(2-2\eps)} (4\eps-2) \,. 
    \end{split}
\end{align}

\subsection{The \texorpdfstring{$b\to s\gamma g$}{b -> s gamma g} amplitude at NLO}

\begin{figure}[t]
    \centering
    \includegraphics[width=0.75\textwidth]{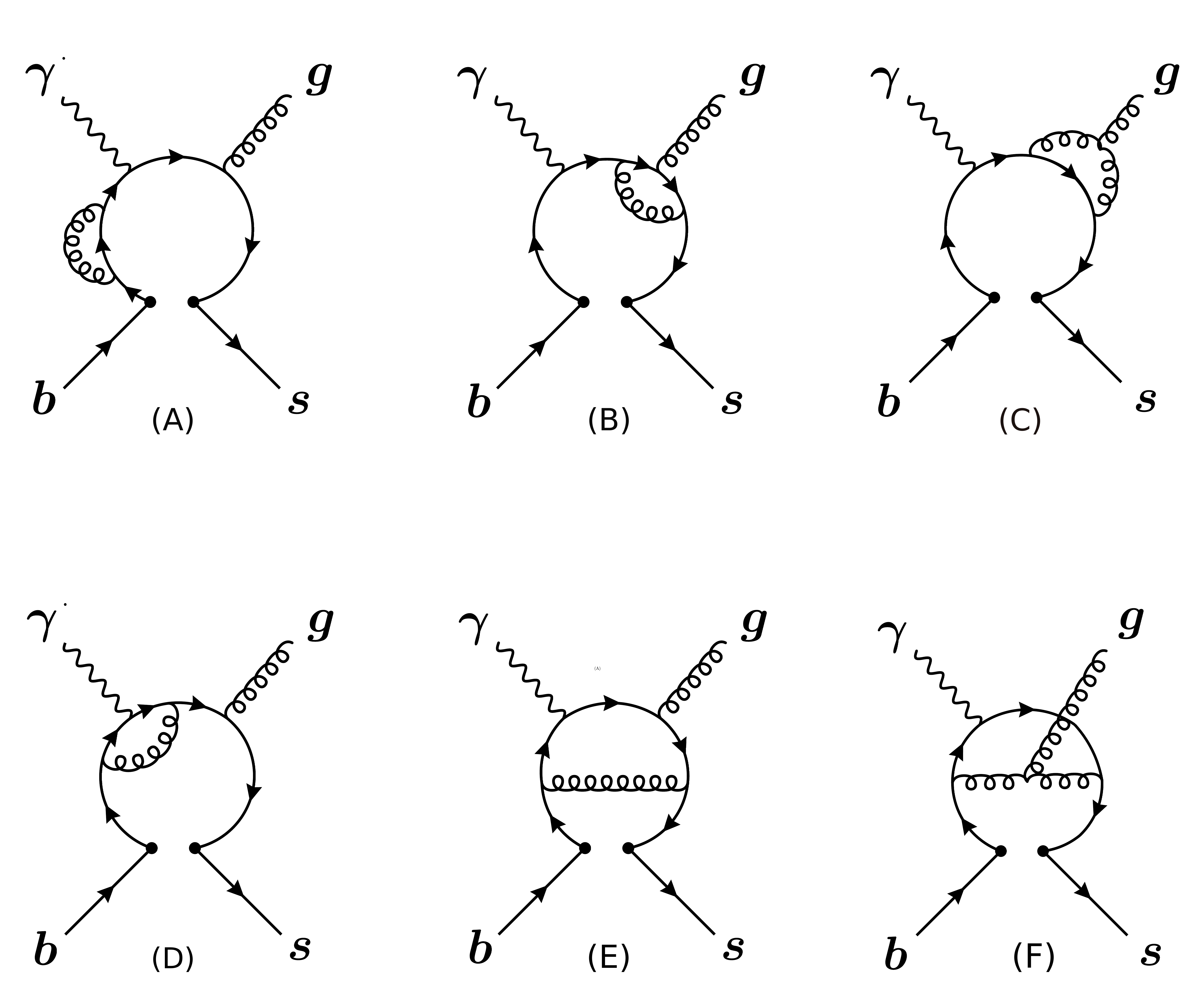}
    \caption{Two-loop diagrams contributing to the $b \to s\gamma g$ amplitude. Only topologies without connections to the external $b$ and $s$ quark lines are shown. Diagram (A) displays one possible insertion of the self-energy correction, but two additional insertions are also possible. Furthermore, for diagrams (B), (C) and (D) the QCD correction to the gluon or photon vertex can be interchanged with the other emitted boson. For diagrams (A) and (E) an additional one can be obtained by exchanging the gluon and photon emission vertices.}
    \label{fig:antijet_NLO_diagrams_1}
\end{figure}

\begin{figure}[t]
    \centering
    \includegraphics[width=0.75\textwidth]{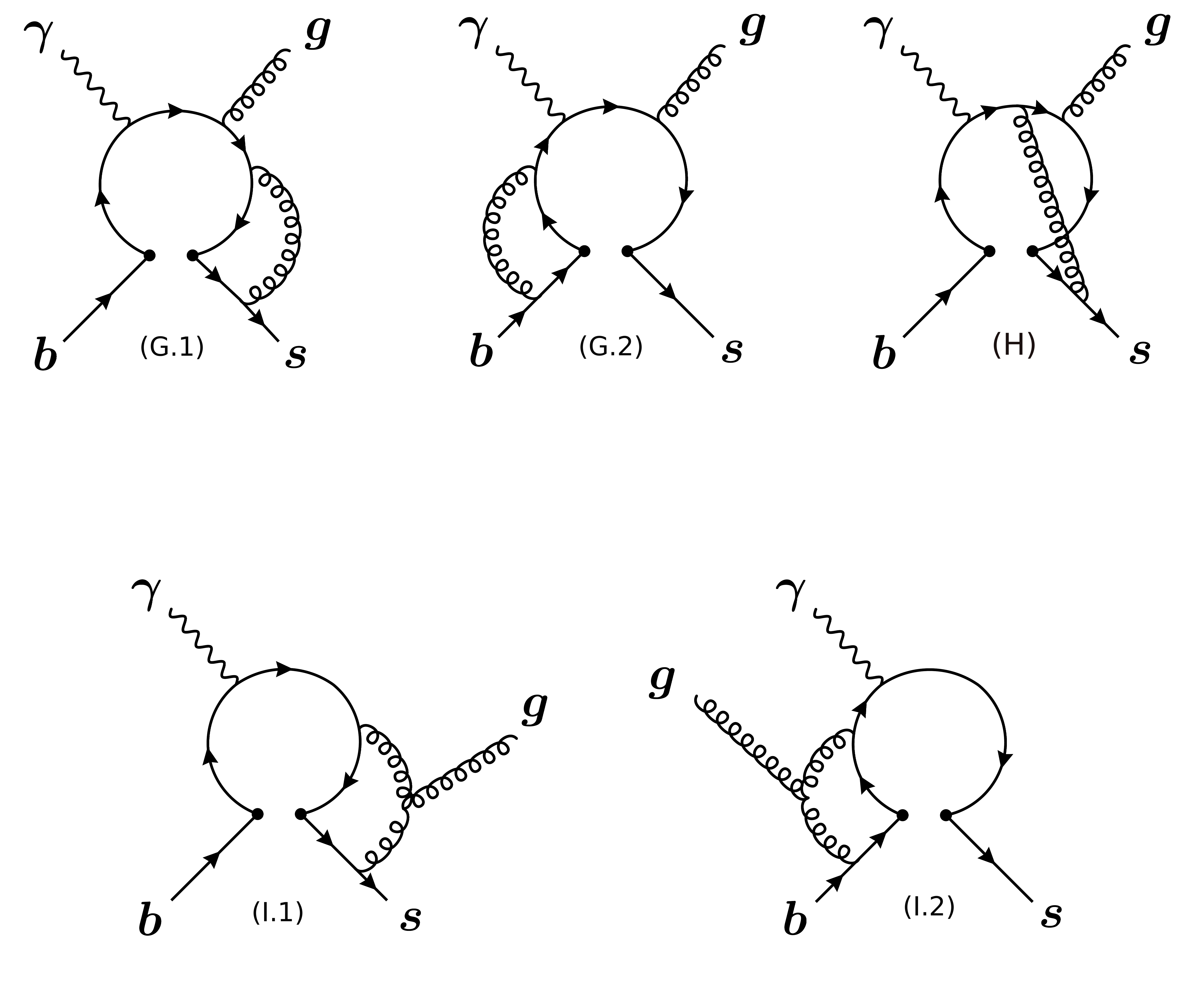}
    \caption{Two-loop diagrams contributing to the $b \to s\gamma g$ amplitude, showing the connections to the external $b$ and $s$ quark lines. In each diagram, the gluon may alternatively be attached to the opposite external leg relative to the one shown. Moreover, for each diagram, an additional one is obtained by interchanging the gluon and photon emission vertices.}
    \label{fig:antijet_NLO_diagrams_2}
\end{figure}

We now proceed to the computation of the $b\to s\gamma g$ amplitude in the anti-hard-collinear region at NLO in the coupling $\alpha_s$. This amounts to evaluating all gluon loop corrections to the LO diagrams shown in Figure~\ref{fig:antijet_LO_diagrams}. We show in Figure~\ref{fig:antijet_NLO_diagrams_1} the two-loop diagrams in which the gluon corrections connect distinct points within the quark loop, and in Figure~\ref{fig:antijet_NLO_diagrams_2} those where the gluon corrections connect the quark loop to the external $b$ and 
$s$ legs.

The reason for separating the diagrams into two categories is that, for those in Figure~\ref{fig:antijet_NLO_diagrams_1}, only the anti-hard-collinear region contributes. In contrast, for those in Figure~\ref{fig:antijet_NLO_diagrams_2}, within our method-of-regions approach, it is necessary to expand the fermion propagator of the external legs, as we will show in the following. 

For diagram (A) we show only one possible insertion of the self-energy correction; however, two additional insertions are also possible: one in the fermion line connecting the photon and the gluon, and one in the fermion line connecting the gluon to the four-quark vertex. In addition, for diagrams (B), (C) and (D) the QCD correction to the gluon or photon vertex can be interchanged with the other emitted boson,  while for diagrams (A) and (E) an additional one can be obtained by exchanging the gluon and photon emission vertices. For diagrams (G), (H), and (I), the gluon may alternatively be attached to the opposite external leg relative to the one shown. Finally, for each diagram in Figure~\ref{fig:antijet_NLO_diagrams_2} an additional one can be obtained by exchanging the gluon and photon emission vertices.  

In this work we restrict ourselves to the $u$-integrated NLO amplitude, which is sufficient to establish the cancellation of all poles. The complete framework also requires the full dependence on the momentum fraction $u$, whose computation we defer to future work. The $u$-integrated result presented here nevertheless constitutes an essential step in this direction, since it will provide the boundary conditions for the differential equations of the master integrals in the $u$-dependent case~\cite{toappearRB}.
Since the present massless computation involves a single scale, the overall momentum dependence factorises completely, in the same way as shown in Equation~\eqref{eq:one_loop_massless_udep}. 

Although the diagrams in Figure~\ref{fig:antijet_NLO_diagrams_2} require an expansion in the anti-hard-collinear region, 
their computation is simpler than that of the diagrams in Figure~\ref{fig:antijet_NLO_diagrams_1}. 
We therefore begin with this class of contributions. 

\begin{figure}[t]
    \centering
    \includegraphics[width=0.90\textwidth]{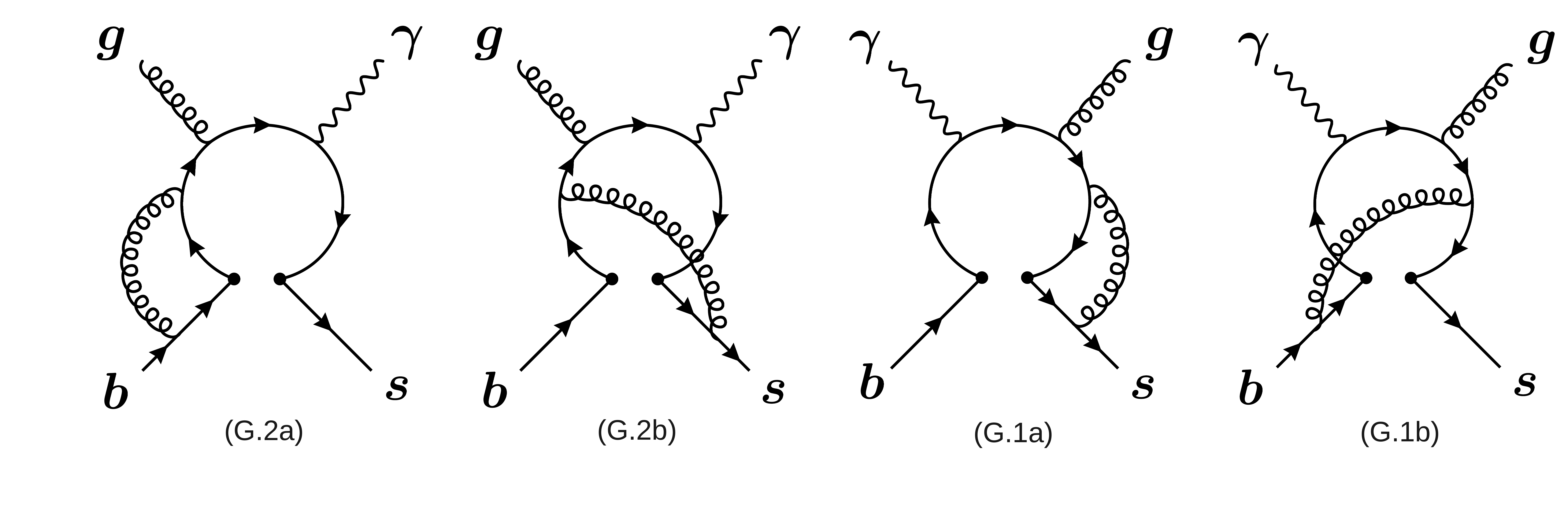}
    \caption{Subset of four diagrams of the kind (G.1) and (G.2).}
    \label{fig:G.1_diagrams}
\end{figure}

In order to illustrate explicitly how the expansion is performed and to highlight the main features of the result of these diagrams, 
we first consider the topology (G.2). 
Including all possible gluon attachments, as well as the interchange of the external photon and gluon, 
this topology gives rise to four distinct diagrams. 
In the following, we focus in particular on the diagrams displayed in Figure~\ref{fig:G.1_diagrams}.


The amplitude for diagram (G.2a) is given by
\begin{align}
\begin{split}
    \mathcal{A}_{\text{NLO}}^{\text{(G.2a)}}=\Tilde{\mu}^{4\eps}\int& \frac{d^Dk}{(2\pi)^D}\int \frac{d^Dl}{(2\pi)^D}\Bar{u}^{(s)}_{hc,L}\gamma^\mu i\frac{\slashed{k}-\slashed{k}_\gamma}{(k-k_\gamma)^2+i0}iQ_u\slashed{\epsilon}_{\perp}^*(k_\gamma) i\frac{\slashed{k}}{k^2+i0}\\
        &\times ig_s\slashed{\epsilon}_{s,\perp}^*(k_g)i\frac{\slashed{k}+\slashed{k}_g}{(k+k_g)^2+i0}ig_st^a\gamma^\alpha i\frac{\slashed{k}+\slashed{k}_g+\slashed{l}}{(k+k_g+l)^2+i0}\\
        &\times\gamma_\mu P_L i\frac{\slashed{k}_b+\slashed{l}+m_b}{(k_b+l)^2-m_b^2+i0}ig_st^a\gamma_\alpha\frac{-i}{l^2+i0} b_L\,,
        \end{split}
\end{align}
where $k_b$ is the incoming $b$-quark momentum. Owing to its connection to the external quark line, this integral is clearly not restricted to the anti-hard-collinear region. To isolate the region of interest, we expand the amplitude and obtain
\begin{align}
\begin{split}
    \mathcal{A}_{\text{NLO}}^{\text{(G.2a)}}=2&g_s^3Q_uC_F\Tilde{\mu}^{4\eps}\int \frac{d^Dk}{(2\pi)^D}\int \frac{d^Dl}{(2\pi)^D}\Bar{u}^{(s)}_{hc,L}\gamma^\mu \frac{\slashed{k}-\slashed{k}_\gamma}{(k-k_\gamma)^2+i0}\slashed{\epsilon}_{\perp}^*(k_\gamma) \frac{\slashed{k}}{k^2+i0}\\
        &\times \slashed{\epsilon}_{s,\perp}^*(k_g)\frac{\slashed{k}}{(k+k_g)^2+i0}\gamma_\mu\frac{1}{(k+k_g+l)^2+i0}\frac{1}{l^2+i0} \frac{l_++k_+}{l_++i0}b_L\,,
        \end{split}
\end{align}
where, in the second line, a linear propagator in the loop momentum $l$ appears in the denominator.

Repeating the same procedure for diagram (G.2b), after the expansion we find
\begin{align}
\begin{split}
    \mathcal{A}_{\text{NLO}}^{\text{(G.2b)}}=-2&g_s^3Q_u\left(C_F-\frac{1}{2}C_A\right)\Tilde{\mu}^{4\eps}\int \frac{d^Dk}{(2\pi)^D}\int \frac{d^Dl}{(2\pi)^D}\Bar{u}^{(s)}_{hc,L}\gamma^\mu \frac{\slashed{k}-\slashed{k}_\gamma}{(k-k_\gamma)^2+i0}\\
&\times\slashed{\epsilon}_{\perp}^*(k_\gamma) \frac{\slashed{k}}{k^2+i0} \slashed{\epsilon}_{s,\perp}^*(k_g)\frac{\slashed{k}}{(k+k_g)^2+i0}\gamma_\mu\frac{1}{(k+k_g+l)^2+i0}\\
&\times\frac{1}{l^2+i0} \frac{l_++k_+}{l_++i0}b_L\,.
        \end{split}
\end{align}
Summing the two amplitudes shows that the colour factor $C_F$ cancels, leaving only $C_A$. This behaviour is common to all amplitudes in Figure~\ref{fig:antijet_NLO_diagrams_2}, which are therefore independent of the colour factor $C_F$.

We now proceed with the standard techniques to evaluate all the diagrams. 
We first perform the tensor reduction and simplify the Dirac structures using the Mathematica package \texttt{FeynCalc}~\cite{Shtabovenko:2016sxi,Shtabovenko:2020gxv,Shtabovenko:2023idz,Mertig:1990an}, 
employing its development version \texttt{10.1.0}. 

After the Dirac algebra has been reduced, the remaining loop integrals involve scalar numerators over a given set of propagators. 
These integrals can be systematically reduced by means of integration-by-parts (IBP) identities~\cite{Chetyrkin:1981qh,Tkachov:1981wb}, 
which exploit total derivative relations in dimensional regularisation to relate different integrals. 
In this way, all integrals can be expressed in terms of a finite set of linearly independent master integrals. 

The IBP reduction is carried out using the package \texttt{FIRE6}~\cite{Smirnov:2019qkx}, 
supplemented by \texttt{LiteRed}~\cite{Lee:2012cn} in order to detect additional symmetries. 
Once the amplitude has been expressed in terms of the resulting minimal set of master integrals, 
these can be evaluated by direct integration or by means of the differential equation method~\cite{Kotikov:1990kg,Kotikov:1991hm,Gehrmann:1999as,Henn:2013pwa}.

The two-loop $u$-independent master integrals belong to a single topology, which we define as
\begin{align}
\label{eq:topA}
\begin{split}
    A = \Bigg\{& \frac{1}{l^2 + i0},\,\frac{1}{k^2 + i0},\, \frac{1}{(k + l)^2 + i0},\,\frac{1}{(l - k_g)^2 + i0}, \,\frac{1}{(k - k_\gamma)^2 + i0},\\
    &\,\frac{1}{(k +l- k_\gamma-k_g)^2 + i0}, \,\frac{1}{k\cdot k_g + i0}\Bigg\} \,.
\end{split}
\end{align}
 In this work, we define the integrals belonging to a generic topology $T$ as
\[
G^{T}_{a_1\ldots a_n} =
\int \frac{d^D l\, d^D k}{(2\pi)^{2D}}
\prod_{i=1}^{n} D_i^{a_i},
\]
where $D_i$ denote the ordered propagators listed in the topology.
  The amplitudes (G), (H), and (I) depend on a single master integral only, namely $G^{A}_{0011100}$. 
Interestingly, this integral does not involve any linear propagator. 

The two-loop master integrals associated with Figure~\ref{fig:antijet_NLO_diagrams_1} are listed in Table~\ref{tab:vc_masters_nonu} for each individual diagram.
\begin{table}[tbhp]
\centering
\begin{tabular}{@{}ll@{}}
\toprule
\textbf{Diagrams} & \textbf{Master integrals} \\
\midrule
(A) & $G^{A}_{0011100},\,G^{A}_{0111010}$ \\
\midrule
(B) & $G^{A}_{0011100},\,G^{A}_{0111010}$ \\
\midrule
(C) & $G^{A}_{0011100},\,G^{A}_{0111010}$ \\
\midrule
(D) & $G^{A}_{0011100},\,G^{A}_{0111010}$ \\
\midrule
(E) & $G^{A}_{0011100},\,G^{A}_{0111010}$ \\
\midrule
(F) & $G^{A}_{0011100},\,G^{A}_{0111010},\,G^{A}_{1111110}$ \\
\bottomrule
\end{tabular}
\caption{Master integrals associated with the diagrams of Figure~\ref{fig:antijet_NLO_diagrams_1}.}
\label{tab:vc_masters_nonu}
\end{table}

All master integrals can be evaluated with standard techniques. Their explicit expressions are collected in Appendix~\ref{app:u-indep-results}, together with the results for the amplitude of each individual diagram. As discussed above, the $u$-integrated results also provide the boundary conditions for the differential equations that will be needed in the $u$-dependent case.

Summing all diagrams, we find that every contribution is proportional to the Dirac structure of the LO amplitude (see Appendix~\ref{app:u-indep-results}). The NLO amplitude can therefore be written as a scalar correction factor multiplying the LO amplitude,
\begin{equation}
\label{eq:NLO_exp}
\mathcal{A}_{\text{NLO}}(b\to s\gamma g) = K(\omega_1;\mu)\; \mathcal{A}_{\text{LO}}(b\to s\gamma g)\,,
\end{equation}
where the LO amplitude~\eqref{eq:one_loop_massless_result_finite} is kept exact in $D=4-2\eps$ dimensions, whereas the correction factor is expanded in $\eps$,
\begin{align}
\label{eq:K_factor}
\begin{split}
K(\omega_1;\mu) = \frac{\alpha_s}{4\pi}\Bigg\{&\Bigg[\frac{2}{\eps^2}+\frac{11}{2\eps}-\frac{2}{\eps}\ln\left(-\frac{2E_\gamma\omega_1}{\mu^2}\right)
+ \ln^2\left( -\frac{2 E_\gamma\, \omega_1}{\mu^2} \right)
- \frac{11}{2} \ln\left( -\frac{2 E_\gamma\, \omega_1}{\mu^2} \right) \\
&- \frac{\pi^2}{2} + \frac{69}{4}\Bigg]C_A + 4C_F + \mathcal{O}(\eps)\Bigg\}\,.
\end{split}
\end{align}
The divergences appear only in the colour structure $C_A$, as a double pole and a single pole with a logarithmic dependence on $E_\gamma\omega_1/\mu^2$, whereas the $C_F$ contribution is finite.
 
Since $K$ carries no Dirac structure, the matching condition~\eqref{antijetmatch} implies that both coefficients $J^{(\bar n),a}$ and $J^{(\bar n),b}$ receive the same correction factor. Through~\eqref{antijet2}, the same therefore holds for the $u$-integrated anti-hard-collinear jet function,
\begin{equation}
\label{eq:NLO_total_u}
J_{\text{NLO}}^{(\bar{n})}(\omega_1) = K(\omega_1;\mu)\; J_{\text{LO}}^{(\bar{n})}(\omega_1)\,,
\end{equation}
with $J_{\text{LO}}^{(\bar{n})}$ given in~\eqref{eq:antijet_LO_u_integrated} after $u$-integration and kept exact in $D$ dimensions. The divergences of $K$ are removed by the multiplicative counterterm
\begin{equation}
\label{eq:Z_antijet}
Z_{\bar J}(\omega_1;\mu) = 1 - \frac{\alpha_s C_A}{4\pi}\left[\frac{2}{\eps^2}+\frac{1}{\eps}\left(\frac{11}{2}-2\ln\left(-\frac{2E_\gamma\omega_1}{\mu^2}\right)\right)\right] + \mathcal{O}(\alpha_s^2)\,,
\end{equation}
which enters the pole cancellation in Section~\ref{sec:pole_cancellation}. For completeness, expanding the product~\eqref{eq:NLO_total_u} in $\eps$ gives
\begin{equation}
\label{eq:NLO_total_u_expanded}
J_{\text{NLO}}^{(\bar{n})}(\omega_1;\,\mu)=\frac{2\alpha_sE_\gamma}{(4\pi)^3}\left(\frac{-2E_\gamma\omega_1}{\mu^2}\right)^{-2\eps}\frac{4\eps -2}{\omega_1+i0}\left(-\frac{2C_A}{\eps^2}-\frac{27C_A}{2\eps}+\left(\frac{2}{3} \pi ^2-\frac{237}{4}\right) C_A-4 C_F\right) + \mathcal{O}(\eps)\,.
\end{equation}
The coefficients of the single pole and of the finite term differ from those in~\eqref{eq:K_factor} only because the $\mathcal{O}(\eps)$ and $\mathcal{O}(\eps^2)$ terms of $J_{\text{LO}}^{(\bar n)}$ multiply the poles of $K$. The genuine divergences are those of $K$.

\section{Cancellation of singularities}
\label{sec:pole_cancellation}
 
Having determined the $\mathcal{O}(\alpha_s)$ corrections to the soft, hard and jet functions entering the factorisation formula~\eqref{FactorisationFormula}, we now demonstrate explicitly the cancellation of divergences of the bare expressions. This provides a non-trivial check of both our results and the underlying factorisation structure of the decay rate.

The UV divergences of the hard corrections are cancelled by the well-known renormalisation of the weak operators and we thus begin by discussing their IR divergences.
In the following we collect the pole terms of the renormalisation factors, which are equal and opposite to the divergences of the corresponding bare functions.

Using the $Z$ factor defined in~\eqref{eq:Z_h}, we extract the divergent part by convolving it with the remaining functions evaluated at LO,
\begin{align}
     \int_0^1 \! du \, Z_{h}(u;\mu) \int_{-\infty}^{\infty} d\omega_1 \, J^{(\bar{n}),{(0)}}(u,\omega_1;\mu) \int_{-\infty}^{\bar{\Lambda}} \! d\omega \, J^{(n),{(0)}}(\omega;\mu) \, g_{17}^{(0)}(\omega,\omega_1;\mu)\,,
\end{align}
where $J^{(\bar{n}),{(0)}}(u,\omega_1;\mu)\equiv J^{(\bar n)}_{\rm LO}(u,\omega_1;\mu)$ is the LO (one-loop) $u$-dependent anti-hard-collinear jet function from Equation~\eqref{eq:one_loop_udep_antijet}, $J^{(n),{(0)}}(\omega;\mu)=\delta(\omega+m_b-2E_\gamma)$ is the LO hard-collinear jet function, and $g_{17}^{(0)}(\omega,\omega_1;\mu)$ is the tree-level subleading shape function.
Performing the $u$-integration, we obtain the following divergent contribution
\begin{align}
\label{eq:hard_poles_convoluted}
    \begin{split}
      \frac{\alpha_s}{4\pi}\int_{-\infty}^{\infty} &d\omega_1 \, J^{(\bar{n}),{(0)}}(\omega_1;\mu) \int_{-\infty}^{\bar{\Lambda}} \! d\omega \, J^{(n),{(0)}}(\omega;\mu) \, g_{17}^{(0)}(\omega,\omega_1;\mu)\\
      &\times\Bigg\{C_F\left[\frac{2}{\eps^2}+\frac{1}{\eps}\left(5+2\ln{\frac{\mu^2}{m_b^2}}\right)\right] +C_A\Bigg[\frac{1}{\eps^2}+\frac{1}{\eps}\left(\ln\frac{\mu^2}{m_b^2}+\frac{7}{2}+i\pi\right)\Bigg]\Bigg\}\,,
      \end{split}
\end{align}
where $J^{(\bar{n}),{(0)}}(\omega_1;\mu)$ denotes the $u$-integrated one-loop anti-hard-collinear jet function~\eqref{eq:antijet_LO_u_integrated}.  

We next consider the hard-collinear jet function, whose renormalisation factor is given in~\eqref{eq:Z-jet}. This yields
\begin{align}
    \label{eq:jet_poles_convoluted}
    \begin{split}
     &-\frac{\alpha_s}{4\pi}C_F\int_{-\infty}^{\infty} d\omega_1 \, J^{(\bar{n}),{(0)}}(\omega_1;\mu) \int_{-\infty}^{\bar{\Lambda}} \! d\omega \, J^{(n),{(0)}}(\omega;\mu) \, g_{17}^{(0)}(\omega,\omega_1;\mu) \left[
    \frac{4}{\eps^2} + \frac{1}{\eps}\left(3 -4\ln\frac{m_b(\bar{\Lambda}-\omega)}{\mu^2}\right)\,
    \right]\\
        &+\frac{\alpha_s}{\pi}\frac{C_F}{\eps}\int_{-\infty}^{\infty} d\omega_1 \, J^{(\bar{n}),{(0)}}(\omega_1;\mu) \int_{0}^{\bar{\Lambda}+m_b-2E_\gamma} \! \frac{d\omega}{\omega} \,   \left(g_{17}^{(0)}(\omega-m_b+2E_\gamma,\omega_1;\mu)-g_{17}^{(0)}(2E_\gamma-m_b,\omega_1;\mu)\right)\,.
    \end{split}
\end{align}
 
The second line contains the non-local contribution arising from the convolution with the star-distribution, which will be crucial for the cancellation of singularities.

Regarding the NLO anti-hard-collinear jet function, since the hard function is taken at leading order and therefore does not depend on $u$, the $u$-integration then acts only on the anti-hard-collinear jet function, so only its $u$-integrated version given in~\eqref{eq:NLO_total_u_expanded} is needed with the explicit poles written in~\eqref{eq:Z_antijet}. We obtain
\begin{align}
    \label{eq:antijet_poles_convoluted}
    \begin{split}
      -\frac{\alpha_s}{4\pi}C_A\int_{-\infty}^{\infty} &d\omega_1 \, J^{(\bar{n}),{(0)}}(\omega_1;\mu) \int_{-\infty}^{\bar{\Lambda}} \! d\omega \, J^{(n),{(0)}}(\omega;\mu) \, g_{17}^{(0)}(\omega,\omega_1;\mu)
      \left[
    \frac{2}{\eps^2} +\frac{1}{\eps} \left(\frac{11}{2} +2\ln\frac{-\mu^2}{2E_\gamma\omega_1}\,
    \right)\right]\,,
    \end{split}
\end{align}
where we note that the singularities depend only on the colour factor $C_A$. We recall that at leading power $2E_\gamma=m_b$ holds.

The remaining contribution arises from the renormalisation of the shape function $g_{17}$, whose $Z$ factor is given in~\eqref{eq:Z-g17}.
To evaluate the convolution of the corresponding distributions, we employ the identities collected in Appendix~\ref{app:distribution_definition}, obtaining
\begin{align}
    \label{eq:g17_poles_convoluted}
    \begin{split}
      &\frac{\alpha_s}{4\pi}\int_{-\infty}^{\infty} d\omega_1 \, J^{(\bar{n}),{(0)}}(\omega_1;\mu) \int_{-\infty}^{\bar{\Lambda}} \! d\omega \, J^{(n),{(0)}}(\omega;\mu) \, g_{17}^{(0)}(\omega,\omega_1;\mu)\\
      &\times\Bigg\{
     C_F \left[
    \frac{2}{\eps^2} - \frac{1}{\eps}\left(2 -4\ln\frac{\mu}{\bar{\Lambda}-\omega}\,
    \right)\right]+
     C_A \left[
    \frac{1}{\eps^2} + \frac{1}{\eps}\left(2 +\ln\frac{\mu^2}{\omega_1^2}+i \pi-2\pi i \theta(-\omega_1)\right)\,
    \right]\Bigg\}\\
        &-\frac{\alpha_s}{\pi}\frac{C_F}{\eps}\int_{-\infty}^{\infty} d\omega_1 \, J^{(\bar{n}),{(0)}}(\omega_1;\mu) \int_{0}^{\bar{\Lambda}+m_b-2E_\gamma} \! \frac{d\omega}{\omega} \,   \left(g_{17}^{(0)}(\omega-m_b+2E_\gamma,\omega_1;\mu)-g_{17}^{(0)}(2E_\gamma-m_b,\omega_1;\mu)\right)\,.
    \end{split}
\end{align}

Finally, combining the contributions in~\eqref{eq:hard_poles_convoluted}, \eqref{eq:jet_poles_convoluted}, \eqref{eq:antijet_poles_convoluted}, and \eqref{eq:g17_poles_convoluted}, we find that all singularities cancel. Equivalently, the renormalisation factors of the individual functions satisfy $Z_h\otimes Z_J\otimes Z_{\bar J}\otimes Z_{17}=1+\mathcal{O}(\alpha_s^2)$, i.e.\ the decay rate is independent of the factorisation scale $\mu$ at this order. This result represents a non-trivial check of the factorisation formula and of all our NLO results.

\section{Cancellation of singularities for the exclusive \texorpdfstring{$\bar{B}_{d,s} \to \gamma \gamma$}{B(d,s) -> gamma gamma} decay}
\label{sec:pole_cancellation_exclusive}
In Refs.~\cite{Bartocci:2024bbf,Bartocci:2025lbl}, we found an important feature of multi-light-cone operators at NLO, for both inclusive and exclusive modes: their anomalous dimensions contain terms that are physically irrelevant. These terms mix the dependences on the different light-cone directions, which would call the factorisation into question.  This feature allows us to simplify the renormalisation  of HQET operators that are composed of fields along distinct light-cone directions.   We have demonstrated this particular feature  explicitly in Refs.~\cite{Bartocci:2024bbf,Bartocci:2025lbl} for the subleading  shape function $g_{17}$. The proof of pole cancellation in the previous section is an important test of this property. We now use the results of Section~\ref{sec:antijet_function} to demonstrate the cancellation of singularities for the exclusive $\bar{B}_{d,s} \to \gamma \gamma$ decay~\cite{Qin:2022rlk}, thereby providing, also in this case, a non-trivial cross-check.  

We emphasise that this result relies crucially on the observation of~\cite{Bartocci:2024bbf, Beneke:2022msp} that
\begin{equation}
    \frac{1}{\omega - \omega' - i0}
    =
    H_+(\omega,\omega')
    -
    H_-(\omega,\omega') \,.
\end{equation}
Since the jet functions relevant for the exclusive $\bar{B}_{d,s} \to \gamma \gamma$ decay possess singularities or branch cuts only in the upper half-plane (i.e. they depend on $\omega-i0$ and $\omega_1-i0$, respectively, where for the exclusive decay we adopt the sign convention for $\omega_1$ of Ref.~\cite{Qin:2022rlk}), one can consistently identify the two distributions,
\begin{equation}
    H_+(\omega,\omega') = H_-(\omega,\omega')\,.
\end{equation}
This identification considerably simplifies the RG evolution and eliminates potentially dangerous mixing terms between the two light-cone directions that could otherwise spoil factorisation~\cite{Bartocci:2024bbf}. The pole cancellation demonstrated in this section shows that this simplification removes only unphysical contributions: the poles associated with the mixing terms between $\omega$ and $\omega_1$ are not needed for the cancellation of singularities.

We assume that the decay amplitude obeys an analogous factorisation formula, with the same convolution structure as in Equation~\eqref{FactorisationFormula} but at the amplitude level, and with $\bar{\Lambda}\to\infty$. The different functions in the factorisation theorem closely resemble the inclusive case. The hard function is now associated exclusively with the operator $Q_1^q$, as derived in Equation~\eqref{eq:Z-factor_hard_Q1}, while the anti-hard-collinear jet function is unchanged up to the replacement $\omega_1\to-\omega_1$. The only non-trivial modification arises in the hard-collinear jet function, whose singular contribution in Equation~\eqref{eq:jet_poles_convoluted} is replaced by
\begin{align}
    \begin{split}
     &-\frac{\alpha_s C_F}{4\pi}\int_{-\infty}^{\infty} d\omega_1 \, J^{(\bar{n}),{(0)}}(\omega_1;\mu) \int_{-\infty}^{\infty} \! d\omega \, J^{(n),{(0)}}(\omega;\mu)  \, \Phi_G^{(0)}(\omega,\omega_1;\mu)
    \left(
        \frac{2}{\eps^2}
        -\frac{2}{\eps}
        \ln\frac{p_-(\omega-i0)}{\mu^2}
    \right)\,,
     \end{split}
\end{align}
where $p_-=m_b$ and the soft function relevant for this process is the $B$-meson soft function $\Phi_G$ defined in~\cite{Qin:2022rlk}.
Its renormalisation properties have been studied in~\cite{Huang:2023jdu,Bartocci:2024bbf}, and inserting these results into the factorisation formula yields the poles\footnote{We remind the reader that in the definition of $\Phi_G$ in~\cite{Huang:2023jdu} one has to take into account the decay constant ${\cal F}_B(\mu)$.}
\begin{align}
    \begin{split}
        &\frac{\alpha_s}{4\pi}\int_{-\infty}^{\infty} d\omega_1 \, J^{(\bar{n}),{(0)}}(\omega_1;\mu) \int_{-\infty}^{\infty} \! d\omega \, J^{(n),{(0)}}(\omega;\mu)  \, \Phi_G^{(0)}(\omega,\omega_1;\mu)
        \\&\times\Bigg\{
        \,C_F
        \left[
            \frac{1}{\eps^2}+\frac{2}{\eps}\ln\frac{\mu}{\omega-i0}-\frac{5}{2\eps}
        \right]+
        C_A
        \left[
            \frac{1}{\eps^2}
            +\frac{2}{\eps}\ln\frac{\mu}{\omega_1-i0}
            +\frac{2}{\eps}
            -\frac{i\pi}{\eps}
        \right]
        \Bigg\}\,.
    \end{split}
\end{align}

Summing all contributions, we find once again that all singularities cancel exactly.

\section{Conclusions}
\label{sec:conclusion}

The resolved contribution arising from the interference of the operators $Q_1^q$ and $Q_{7\gamma}$ is the numerically dominant subleading-power correction to the inclusive $\bar B\to X_s\gamma$ decay and one of the main sources of its theoretical uncertainty. Previous analyses at lowest order revealed a very large scale ambiguity and a pronounced dependence on the charm-quark mass. Both call for a systematic treatment of $\alpha_s$ corrections within a factorisation framework. In this work we have taken a further step in this direction at next-to-leading order.

We have presented a formulation of the factorisation theorem for the resolved $Q_1^q$–$Q_{7\gamma}$ contribution that is valid beyond leading order. It separates the rate into a hard function, the hard-collinear and anti-hard-collinear jet functions, and the subleading shape function $g_{17}$. By means of a dedicated matching calculation, we have shown that the anti-hard-collinear jet function can be uniquely extracted, to all orders in $\alpha_s$, from the anti-hard-collinear region of the $b\to s\gamma g$ amplitude induced by $Q_1^q$. We have verified at leading order that this definition agrees with the one arising directly from the factorisation formula. We then addressed the NLO corrections to the functions entering the factorisation formula, with the aim of demonstrating the cancellation of all poles in their sum. This provides a non-trivial check of the calculation and, at the same time, validates the consistency of the factorisation framework.   To this end, we computed the one-loop hard-matching corrections and extracted the infrared poles required to demonstrate their cancellation; the fully renormalised results including the evanescent contributions will be presented in~\cite{toappearRB}.

An important technical result of this paper is the two-loop anti-hard-collinear jet function for a massless quark loop, which is not only essential for establishing the cancellation of poles, but also provides the basis for the extensions discussed below.

Combining these ingredients with the known hard-collinear jet function and with the renormalisation of $g_{17}$ derived in our previous work~\cite{Bartocci:2024bbf}, we have demonstrated explicitly that all $1/\eps$ singularities cancel between the hard, hard-collinear, anti-hard-collinear and soft sectors. This provides a non-trivial consistency check of the factorisation framework and of the individual NLO results. It also confirms that the physically irrelevant terms previously identified in the anomalous dimension of the multi-light-cone soft operator $g_{17}$ can be consistently omitted. We have shown that the same holds for the exclusive decays $\bar B_{d,s}\to\gamma\gamma$.

The present results lay the groundwork for a complete NLO analysis of this resolved contribution. In this work we have restricted ourselves to the $u$-integrated two-loop anti-hard-collinear jet function, which is sufficient to establish the cancellation of all poles. The anti-hard-collinear jet function with its full dependence on the momentum fraction $u$, which is required once the hard function is included beyond leading order, still has to be computed and verified; this is not done in the present paper. The $u$-integrated result obtained here provides the boundary conditions for the differential equations of the master integrals in the $u$-dependent case. A further step is the inclusion of charm-quark mass effects in the two-loop anti-hard-collinear jet function. Since the $1/\eps$ poles do not depend on the charm-quark mass, the pole structure established here provides a stringent check of these calculations. Together with the renormalisation-group evolution of the individual functions, this will allow a resummation of the large logarithms between the hard, hard-collinear and soft scales. The NLO analysis will also reduce the scale ambiguity of the LO result and will establish a well-defined residual scale dependence of $\mathcal{O}(\alpha_s^2)$. It may also reduce the dependence on the charm-quark mass.
Thus, we expect such a phenomenological analysis to lead to a significantly improved estimate of this contribution, and hence to a more precise Standard Model prediction for the $\bar B\to X_s\gamma$ branching ratio.

\section*{Acknowledgments} 
 We thank Philipp Böer for crucial comments and discussions. We thank Michael Benzke, Kevin Brune,  Christoph Greub and Vladyslav Shtabovenko for useful discussions. We thank Kevin Brune for his careful reading of the manuscript. This work is supported by  the  Cluster  of  Excellence  ``Precision  Physics,  Fundamental Interactions, and Structure of Matter" (PRISMA$^{++}$ EXC 2118/2) funded by the German Research Foundation (DFG) within the German Excellence Strategy (Project ID 390831469) and also by the DFG Grant HU 901/2-1.

\begin{appendix}

\newpage
\section{Conventions}
\label{app:setup}

\paragraph{Weak effective Hamiltonian.}
We use the conventions of Ref.~\cite{Beneke:2001ev}. Assuming CKM unitarity, the effective Hamiltonian for $b\to s$ transitions reads
\begin{equation}
\label{eq:Heff}
{\cal H}_{\rm eff} = \frac{G_F}{\sqrt{2}} \sum_{q=u,c} \lambda_q \bigg( C_1^q\, Q_1^q + C_2^q\, Q_2^q + \sum_{i=3}^{6} C_i\, Q_i + C_{7\gamma}\, Q_{7\gamma} + C_{8g}\, Q_{8g} \bigg) + \text{h.c.}\,,
\end{equation}
with $\lambda_q = V_{qb} V_{qs}^*$. The Wilson coefficients do not depend on the flavour $q$ of the current-current operators; we nevertheless keep the superscript in $C_{1,2}^q$ to indicate the operator $Q_{1,2}^q$ they multiply. The operators relevant for our analysis are ($i$ and $j$ are colour indices)
\begin{equation}
\label{eq:operators}
\begin{aligned}
   Q_1^q &= \bar q_i \gamma^\mu (1-\gamma_5) b_i \;\, \bar s_j \gamma_\mu (1-\gamma_5) q_j \,, &\qquad
   Q_2^q &= \bar q_i \gamma^\mu (1-\gamma_5) b_j \;\, \bar s_j \gamma_\mu (1-\gamma_5) q_i \,, \\
   Q_{7\gamma} &= \frac{-e\, m_b}{8\pi^2}\, \bar s \sigma_{\mu\nu} (1+\gamma_5) F^{\mu\nu} b \,, &\qquad
   Q_{8g} &= \frac{-g_s\, m_b}{8\pi^2}\, \bar s \sigma_{\mu\nu} (1+\gamma_5) G^{\mu\nu} b \,,
\end{aligned}
\end{equation}
while the explicit form of the QCD penguin operators $Q_{3,\dots,6}$, which do not enter the present analysis, can be found in~\cite{Beneke:2001ev}. In Section~\ref{sec:factorisation} the Fierz-rearranged form of $Q_1^q$ is used. The sign conventions correspond to the covariant derivative $iD_\mu = i\partial_\mu + g_s t^a A^a_\mu + e\, Q_f A_\mu$, where $t^a$ are the $SU(3)$ colour generators and $Q_f$ is the electric charge of the fermion $f$ in units of $e$ (e.g.\ $Q_u = 2/3$).

\paragraph{Kinematics and power counting.}
We consider the CP-averaged photon energy spectrum in the endpoint region $m_b - 2E_\gamma = \mathcal{O}(\Lambda_{\rm QCD})$. We introduce two light-like reference vectors with $n^2 = \bar n^2 = 0$, $n\cdot\bar n = 2$, and work in the $\bar B$ rest frame, $v^\mu = (n^\mu + \bar n^\mu)/2$. Any four-vector is decomposed as
\begin{equation}
a^\mu = a_+ \frac{\bar n^\mu}{2} + a_- \frac{n^\mu}{2} + a_\perp^\mu \,, \qquad a_+ \equiv n\cdot a\,, \quad a_- \equiv \bar n\cdot a\,,
\end{equation}
such that $a^2 = a_+ a_- + a_\perp^2$. For position arguments along the light cone we write $x_-^\mu \equiv (\bar n\cdot x)\, n^\mu/2$ and $x_+^\mu \equiv (n\cdot x)\, \bar n^\mu/2$. The photon momentum is $q^\mu = E_\gamma \bar n^\mu$, i.e.\ $q_+ = 2E_\gamma$. Writing the $b$-quark momentum as $p_b = m_b v + k$ with a soft residual momentum $k$, the hard-collinear strange-quark jet carries $p = p_b - q$ with $p_- = m_b + \mathcal{O}(\Lambda_{\rm QCD})$ and $p_+ = m_b - 2E_\gamma + k_+$, hence $p^2 = m_b\,(m_b - 2E_\gamma + k_+)$ at leading power.\footnote{Throughout, $p_+$ denotes the complete $+$ component $n\cdot p$ of the momentum $p$. At tree level $p_+ = m_b - 2E_\gamma$, which differs from the hadronic variable $n\cdot P_X = M_B - 2E_\gamma$ by $\bar\Lambda = M_B - m_b$.}

The endpoint region is characterised by three scales: a hard scale $\mathcal{O}(m_b)$, an (anti-)hard-collinear scale $\mathcal{O}(\sqrt{m_b\Lambda_{\rm QCD}})$ and a soft scale $\mathcal{O}(\Lambda_{\rm QCD})$. With the expansion parameter $\lambda^2 = \Lambda_{\rm QCD}/m_b$,\footnote{An alternative convention often used in the literature is $\lambda = \Lambda_{\rm QCD}/m_b$.} the relevant momentum modes scale as
\begin{equation}
(a_+, a_-, a_\perp) \sim
\begin{cases}
(1,1,1)\, m_b & \text{hard}\,, \\
(\lambda^2,1,\lambda)\, m_b & \text{hard-collinear}\,, \\
(1,\lambda^2,\lambda)\, m_b & \text{anti-hard-collinear}\,, \\
(\lambda^2,\lambda^2,\lambda^2)\, m_b & \text{soft}\,.
\end{cases}
\end{equation}
The strange-quark jet is hard-collinear, while the photon and the light-quark loop are anti-hard-collinear. Since the soft modes have parametrically smaller virtuality than the (anti-)hard-collinear ones, the decay is described by $\mathrm{SCET}_{\rm I}$~\cite{Bauer:2000yr,Bauer:2001yt,Beneke:2002ph,Beneke:2002ni} combined with HQET for the heavy quark, $\slashed{v} h_v = h_v$.

\paragraph{Gauge-invariant building blocks.}
Following Ref.~\cite{Hill:2002vw}, the hard-collinear quark and gluon fields enter the effective operators only through the gauge-invariant combinations
\begin{equation}
\chi^i_{hc} = \big(W_{hc}^\dagger\big)^{ij}\, \xi^j_{hc}\,, \qquad
\mathcal{A}^\mu_{hc} = W_{hc}^\dagger \big[ iD^\mu_{hc} W_{hc} \big]\,,
\end{equation}
where $i,j$ are colour indices, $\slashed{n}\,\xi_{hc} = 0$, $A^\mu_{hc} = A^{a\mu}_{hc}\, t^a$, and $iD^\mu_{hc} = i\partial^\mu + g_s A^\mu_{hc}$. Expanding in the coupling, the gluon building block reads
\begin{equation}
\mathcal{A}^\mu_{hc} = g_s\, t^a \Big( A^{a\mu}_{hc} - \frac{\partial^\mu}{\bar n\cdot\partial}\, \bar n\cdot A^a_{hc} \Big) + \mathcal{O}(g_s^2)\,,
\end{equation}
where the non-abelian terms of higher order in $g_s$ are not needed in this work. The Wilson line
\begin{equation}
W_{hc}(x) = \mathrm{P} \exp\bigg[ i g_s \int_{-\infty}^0 ds\; \bar n\cdot A_{hc}(x + s\bar n) \bigg]
\end{equation}
is a matrix in colour space. The anti-hard-collinear building blocks $\chi_{\overline{hc}}$ and $\mathcal{A}^\mu_{\overline{hc}}$ follow from the replacement $n \leftrightarrow \bar n$, with $\slashed{\bar n}\,\xi_{\overline{hc}} = 0$. For the anti-hard-collinear photon, the abelian analogue with $g_s A^{a\mu} t^a \to e A^\mu$ gives exactly
\begin{equation}
\mathcal{A}^{\mu}_{\perp,\overline{hc}} = e \Big( A_\perp^\mu - \frac{\partial_\perp^\mu}{n\cdot\partial}\, n\cdot A \Big)\,.
\end{equation}
All (anti-)hard-collinear fields in this work are understood after soft decoupling, $\xi_{\overline{hc}}\to S_{\bar n}\,\xi_{\overline{hc}}$ (and analogously $\xi_{hc}\to S_n\,\xi_{hc}$), with the soft Wilson lines
\begin{equation}
S_{\bar n}(x) = \mathrm{P}\exp\bigg[ i g_s \int_{-\infty}^0 ds\; \bar n\cdot A_s(x+s\bar n)\bigg]\,,
\end{equation}
and $S_n$ obtained by $\bar n\to n$. The soft gluon then enters through the gauge-invariant building block
\begin{equation}
\label{eq:soft_building_block}
\mathcal{A}^\mu_{s}(x) = S_{\bar n}^\dagger(x)\big[iD_s^\mu S_{\bar n}(x)\big]\,,
\end{equation}
whose relation to the field-strength tensor is given in~\eqref{eq:A_vs_G}.

\section{Definitions of the distributions used in this work}
\label{app:distribution_definition}

In this appendix we collect the definitions of the distributions used throughout the paper.

We begin with the standard plus-distribution, defined through its action on a smooth test function $f(\omega')$ as
\begin{equation}
    \int d\omega'\, [g(\omega,\omega')]_{+}\, f(\omega')
    \equiv
    \int d\omega'\, g(\omega,\omega') \left[f(\omega') - f(\omega)\right] \,.
\end{equation}
If the support does not extend to $+\infty$, we denote this by a superscript $(\bar{\Lambda})$, indicating that the integration over $\omega'$ is restricted to the interval $(-\infty, \bar{\Lambda}]$:
\begin{equation}
    \int_{-\infty}^{\bar{\Lambda}} \! d\omega' \, [g(\omega,\omega')]_{+}^{(\bar{\Lambda})} \, f(\omega') \equiv \int_{-\infty}^{\bar{\Lambda}} \! d\omega' \, g(\omega,\omega') \, (f(\omega')-f(\omega)) \,.
\end{equation}
A useful generalisation of the plus-distribution is provided by the star-distributions, defined as~\cite{DeFazio:1999ptt}
\begin{equation}
\label{eq:star_distributions}
\begin{aligned}
   \int_0^z dx\, f(x) 
   \left( \frac{1}{x} \right)_{\!*}^{[u]}
   &\equiv \int_0^z dx\, \frac{f(x)-f(0)}{x} + f(0)\,\ln\frac{z}{u} \,, \\
   \int_0^z dx\, f(x)
    \left( \frac{\ln{x/u}}{x} \right)_{\!*}^{[u]}
   &\equiv \int_0^z dx\, \frac{f(x)-f(0)}{x}\,\ln\frac{x}{u} 
    + \frac{f(0)}{2}\,\ln^2\frac{z}{u} \,,
\end{aligned}
\end{equation}
where $f(x)$ is assumed to be a smooth test function.
When the integration variables can take both positive and negative values, it is convenient to introduce the modified plus-distributions
\begin{equation}
    \int d\omega'\, [g(\omega,\omega')]_{\oplus/\ominus}\, f(\omega')
    \equiv
    \int d\omega'\, g(\omega,\omega') 
    \left[f(\omega') - \theta(\pm \omega') f(\omega)\right] \,.
\end{equation}
Using these building blocks, we define the distributions
\begin{eqnarray}
\label{eq:FandGdistributions}
F^{>}(\omega, \omega') & = &
\left [ \frac{\omega \, \theta (\omega' - \omega)} {\omega' \,  (\omega' - \omega)}   \right ]_{+}
+ \left [ \frac{\theta (\omega - \omega')}  {\omega - \omega'}  \right ]_{\oplus} \,,
\nonumber \\
F^{<}(\omega, \omega')  & = &
\left [ \frac{\omega \, \theta (\omega - \omega')} {\omega' \,  (\omega- \omega')}   \right ]_{+}
+ \left [ \frac{\theta (\omega' - \omega)}  {\omega' - \omega}  \right ]_{\ominus}  \,,
\nonumber \\
G^{>}(\omega, \omega')  & = & (\omega + \omega') \,
\left [ \frac{\theta (\omega' - \omega)} {\omega' \,  (\omega' - \omega)}   \right ]_{+}
- i \pi \delta(\omega - \omega') \,,
\nonumber \\
G^{<}(\omega, \omega')  & = &   (\omega + \omega') \,
\left [ \frac{ \theta (\omega - \omega')} {\omega' \,  (\omega- \omega')}   \right ]_{+}
+ i \pi \delta(\omega - \omega') \,,
\end{eqnarray}
which have been introduced in the study of QED corrections to the $B$-meson light-cone distribution amplitude in~\cite{Beneke:2022msp}.
Finally, it is convenient to introduce the combinations
\begin{equation}
\label{eq:H_distributions}
H_{\pm}(\omega, \omega') =  
\theta( \pm \omega)\, F^{>(<)} (\omega, \omega')
+ \theta( \mp \omega)\, G^{<(>)} (\omega, \omega') \,,
\end{equation}
 as well as~\cite{Bartocci:2024bbf}
\begin{equation}
    \mathbf{H}(\omega_1,\omega_1') = H_+(\omega_1,\omega_1') + H_-(\omega_1,\omega_1') \,.
\end{equation}
Additionally, we provide in this appendix the following useful distributional identities
\begin{align}
    \begin{split}
   &  \int_{-\infty}^{\infty}d\omega_1 \frac{1}{\omega_1+i0} \int_{-\infty}^{\infty}d\omega_1^\prime\left ( {\omega_1 \over \omega_1^{\prime  \, 2}} \right )\left [ \theta(\omega_1)  \theta(\omega_1^{\prime} - \omega_1)
-  \theta(- \omega_1)  \theta(\omega_1 - \omega_1^{\prime})  \right ]f(\omega_1^\prime)\\
&= \int_{-\infty}^{\infty} d\omega_1 \frac{1}{\omega_1+i0}f(\omega_1)\,,
    \end{split}
\end{align}
and
\begin{align}
    \begin{split}
   &  \int_{-\infty}^{\infty}d\omega_1 \frac{1}{\omega_1+i0} \int_{-\infty}^{\infty}d\omega_1^\prime H_+(\omega_1,\omega_1^\prime)f(\omega_1^\prime)=\int_{-\infty}^{\infty}d\omega_1 \frac{1}{\omega_1+i0}\left(2i\pi\theta(-\omega_1)\right)f(\omega_1)\,.
    \end{split}
\end{align}

\section{Master integrals and diagram-by-diagram results for the \texorpdfstring{$u$}{u}-integrated two-loop amplitude}
\label{app:u-indep-results}
In this appendix we collect the results for the master integrals associated with the topology defined in Equation~\eqref{eq:topA} and listed in Table~\ref{tab:vc_masters_nonu}. 
Their explicit expressions read as follows:

\begin{align}
\label{eq:NLO_G3}
    \begin{split}
        G^{A}_{0011100}=\frac{(-2E_\gamma\omega_1)^{1-2\eps}}{(4\pi)^{4-2\eps}}\frac{\Gamma(1-\eps)^3\Gamma(2\eps-1)}{\Gamma(3-3\eps)}\,,
    \end{split}
\end{align}

\begin{align}
    \begin{split}
        G^{A}_{0111010}=-\frac{(-2E_\gamma\omega_1)^{-2\eps}}{(4\pi)^{4-2\eps}}\frac{\Gamma(\eps)\Gamma(2\eps)\Gamma(1-2\eps)\Gamma(1-\eps)^2}{(1-2\eps)\Gamma(2-3\eps)}\,,
    \end{split}
\end{align}

\begin{align}
    \begin{split}
        G^{A}_{1111110}=-&\frac{(-2E_\gamma\omega_1)^{-2-2\eps}}{(4\pi)^{4-2\eps}}\frac{1}{\Gamma(1 - \eps)^2}\Bigg[
\frac{1}{\eps^4}
- \frac{5 \pi^2}{6\, \eps^2}
- \frac{27\, \zeta(3)}{\eps}
- \frac{23 \pi^4}{36}
\\
&+ \eps \left(8 \pi^2 \zeta(3) - 117 \zeta(5)\right)
- \eps^2 \left(\frac{19 \pi^6}{315} - 267 \zeta(3)^2\right)
\\
&
+ \eps^3 \left(\frac{109 \pi^4 \zeta(3)}{10} + 40 \pi^2 \zeta(5) + 6 \zeta(7)\right)\Bigg]
\,,
    \end{split}
\end{align}
in agreement with~\cite{gehrmann2010calculation}.
Note that, in contrast to Equation~\eqref{eq:one_loop_massless_result_finite}, the results below are written in terms of the commutator $[\slashed{\epsilon}_{\perp}^*(k_\gamma),\slashed{\epsilon}_{s,\perp}^*(k_g)]=-[\slashed{\epsilon}_{s,\perp}^*(k_g),\slashed{\epsilon}_{\perp}^*(k_\gamma)]$, which has to be taken into account when comparing with the LO amplitude.

Using these three master integrals, one obtains the final amplitudes for all diagrams. 
Starting from Figure~\ref{fig:antijet_NLO_diagrams_1}, we find

\begin{align}
    \label{eq:NLO_A_result}
    \begin{split}
         \mathcal{A}_{\text{NLO}}^{\text{(A)}}=&-\frac{4g_s\alpha_sQ_uC_FE_\gamma}{(4\pi)^3} \left(\frac{-2E_\gamma\omega_1}{\mu^2}\right)^{-2\eps}e^{2\gamma_E\eps}\Bar{u}^{(s)}_{hc,L} \left[\slashed{\epsilon}_{\perp}^*(k_\gamma),\slashed{\epsilon}_{s,\perp}^*(k_g)\right]\frac{\slashed{\bar{n}}}{2} b_L \\
         &\times\frac{(1+\eps)\Gamma(1-\eps)\Gamma(2\eps)\left[(3\eps-3)\Gamma(1-\eps)^2-\eps\Gamma(1-2\eps)\Gamma(2-\eps)\Gamma(\eps)\right]}{(2\eps-1)\Gamma(3-3\eps)} \,,
    \end{split}
\end{align}
\begin{align}
    \label{eq:NLO_B_result}
    \begin{split}
         \mathcal{A}_{\text{NLO}}^{\text{(B)}}=&\frac{4g_s\alpha_sQ_u(C_A-2C_F)E_\gamma}{(4\pi)^3} \left(\frac{-2E_\gamma\omega_1}{\mu^2}\right)^{-2\eps}\Bar{u}^{(s)}_{hc,L} \left[\slashed{\epsilon}_{\perp}^*(k_\gamma),\slashed{\epsilon}_{s,\perp}^*(k_g)\right]\frac{\slashed{\bar{n}}}{2} b_L \\
         &\times e^{2\gamma_E\eps}\Big[(1+\eps)\Gamma(1-2\eps)\Gamma(1-\eps)\Gamma(\eps)\Gamma(2\eps)-\Gamma(2-\eps)\Gamma(-\eps)\Gamma(2\eps-1)\Big]\\
         &\times\frac{(1+\eps)\Gamma(2-\eps)}{(2\eps-1)\Gamma(3-3\eps)}\,,
    \end{split}
\end{align}
\vspace{-1em}
\begin{align}
    \label{eq:NLO_C_result}
    \begin{split}
         \mathcal{A}_{\text{NLO}}^{\text{(C)}}=&\frac{2g_s\alpha_sQ_uC_AE_\gamma}{(4\pi)^3} \left(\frac{-2E_\gamma\omega_1}{\mu^2}\right)^{-2\eps}e^{2\gamma_E\eps}\Bar{u}^{(s)}_{hc,L} \left[\slashed{\epsilon}_{\perp}^*(k_\gamma),\slashed{\epsilon}_{s,\perp}^*(k_g)\right]\frac{\slashed{\bar{n}}}{2} b_L \\
         &\times\Big[(3 \eps-1) \Gamma (3-2 \eps) \Gamma (1-\eps) \Gamma (\eps)-2 (\eps (2
   \eps-5)+1) \Gamma (2-\eps) \Gamma (-\eps)\Big]\\
         &\times\frac{(1+\eps)\Gamma(2-\eps)\Gamma(1-2\eps)\Gamma(2\eps)}{\eps(2\eps-1)\Gamma(3-3\eps)\Gamma(3-2\eps)}\,,
    \end{split}
\end{align}
\vspace{-1em}
\begin{align}
    \label{eq:NLO_D_result}
    \begin{split}
         \mathcal{A}_{\text{NLO}}^{\text{(D)}}=&\frac{8g_s\alpha_sQ_uC_FE_\gamma}{(4\pi)^3} \left(\frac{-2E_\gamma\omega_1}{\mu^2}\right)^{-2\eps}e^{2\gamma_E\eps}\Bar{u}^{(s)}_{hc,L} \left[\slashed{\epsilon}_{\perp}^*(k_\gamma),\slashed{\epsilon}_{s,\perp}^*(k_g)\right]\frac{\slashed{\bar{n}}}{2} b_L \\
         &\times\Big[(2\eps^3+\eps^2+\eps-1)\Gamma(1+\eps)\Gamma(1-2\eps)+(3\eps^3-\eps^2+1)\Gamma(1-\eps)\Big]\\
         &\times\frac{(1+\eps)\Gamma(1-\eps)^2\Gamma(2\eps-1)}{\eps(2\eps-1)\Gamma(3-3\eps)}\,,
    \end{split}
\end{align}
\begin{align}
    \label{eq:NLO_E_result}
    \begin{split}
         \mathcal{A}_{\text{NLO}}^{\text{(E)}}=&\frac{g_s\alpha_sQ_u(C_A-2C_F)E_\gamma}{2(4\pi)^3} \left(\frac{-2E_\gamma\omega_1}{\mu^2}\right)^{-2\eps}e^{2\gamma_E\eps}\Bar{u}^{(s)}_{hc,L} \left[\slashed{\epsilon}_{\perp}^*(k_\gamma),\slashed{\epsilon}_{s,\perp}^*(k_g)\right]\frac{\slashed{\bar{n}}}{2} b_L \\
         &\times\Bigg[
\frac{
4 \eps \left(4 \eps^3 + 10 \eps^2 - 11 \eps + 3\right) 
\Gamma(1 - 2 \eps)\, \Gamma(2 \eps)
}{(2 \eps - 1)\, \Gamma(3 - 3 \eps)}
 \\
&\quad
\left.
-
\frac{
4 \left(6 \eps^5 - 11 \eps^4 + 42 \eps^3 - 52 \eps^2 + 25 \eps - 4\right) 
\Gamma(-\eps)\, \Gamma(2 \eps - 1)
}{(2 \eps - 1)\, \Gamma(3 - 3 \eps)\, \Gamma(\eps)}
\right. \\
&\quad
-
\frac{
8 \left(\eps \left(\eps (2 \eps + 3) + 3\right) - 2\right)\, 
\Gamma(2 - \eps)^2\, \Gamma(\eps)
}{\Gamma(3 - 2 \eps)^2}
\Bigg]\times\frac{(1+\eps)\Gamma(1-\eps)^2\Gamma(\eps)}{(1-\eps)^2}\,,
    \end{split}
\end{align}
\begin{align}
    \label{eq:NLO_F_result}
    \begin{split}
         \mathcal{A}_{\text{NLO}}^{\text{(F)}}=&-\frac{g_s\alpha_sQ_uC_AE_\gamma}{(4\pi)^3} \left(\frac{-2E_\gamma\omega_1}{\mu^2}\right)^{-2\eps}\Bar{u}^{(s)}_{hc,L} \left[\slashed{\epsilon}_{\perp}^*(k_\gamma),\slashed{\epsilon}_{s,\perp}^*(k_g)\right]\frac{\slashed{\bar{n}}}{2} b_L\\
         &\times\left[\frac{1}{\eps}\left(1+\frac{\pi^2}{6}\right)+\left(\zeta(3)+\frac{21}{2}+\frac{11}{12}\pi^2\right)+\mathcal{O}(\eps)\right]\,.
    \end{split}
\end{align}
Since the result for diagram (F) is given in expanded form, the factor $e^{2\gamma_E\eps}$ has been absorbed into the coefficients of the $\eps$-expansion. The results for the diagrams in Figure~\ref{fig:antijet_NLO_diagrams_2} are:
\begin{align}
\label{eq:NLO_G12_result}
    \begin{split}
        \mathcal{A}_{\text{NLO}}^{\text{(G.1)}+\text{(G.2)}}=&-\frac{12g_s\alpha_sQ_uC_AE_\gamma}{(4\pi)^3}\left(\frac{-2E_\gamma\omega_1}{\mu^2}\right)^{-2\eps}\Bar{u}^{(s)}_{hc,L} \left[\slashed{\epsilon}_{\perp}^*(k_\gamma),\slashed{\epsilon}_{s,\perp}^*(k_g)\right]\frac{\slashed{\bar{n}}}{2} b_L\\
        &\times e^{2\gamma_E\eps}\frac{(1+\eps)\Gamma(2-\eps)^2\Gamma(-\eps)\Gamma(2\eps)}{\Gamma(3-3\eps)(2\eps-1)(\eps-1)}\,,
    \end{split}
\end{align}

\begin{align}
    \label{eq:NLO_H_result}
    \begin{split}
         \mathcal{A}_{\text{NLO}}^{\text{(H)}}=&-\frac{4g_s\alpha_sQ_uC_AE_\gamma}{(4\pi)^3} \left(\frac{-2E_\gamma\omega_1}{\mu^2}\right)^{-2\eps}\Bar{u}^{(s)}_{hc,L} \left[\slashed{\epsilon}_{\perp}^*(k_\gamma),\slashed{\epsilon}_{s,\perp}^*(k_g)\right]\frac{\slashed{\bar{n}}}{2} b_L\\
        &\times e^{2\gamma_E\eps} \frac{(1+\eps)\Gamma(1-2\eps)\Gamma(-\eps)\Gamma(2-\eps)^2\Gamma(2\eps-1)}{\Gamma(3-3\eps)\Gamma(2-2\eps)(1-\eps)}\,,
    \end{split}
\end{align}
\begin{align}
    \label{eq:NLO_I_result}
    \begin{split}
        \mathcal{A}_{\text{NLO}}^{\text{(I.1)}+\text{(I.2)}}=&-\frac{12g_s\alpha_sQ_uC_AE_\gamma}{(4\pi)^3} \left(\frac{-2E_\gamma\omega_1}{\mu^2}\right)^{-2\eps}\Bar{u}^{(s)}_{hc,L} \left[\slashed{\epsilon}_{\perp}^*(k_\gamma),\slashed{\epsilon}_{s,\perp}^*(k_g)\right]\frac{\slashed{\bar{n}}}{2} b_L\\
        &\times e^{2\gamma_E\eps} \frac{(1+\eps)(\eps-2)\Gamma(2\eps)\Gamma(1-2\eps)\Gamma(2-\eps)^2\Gamma(-\eps)}{(2\eps-1)\Gamma(2-2\eps)\Gamma(4-3\eps)}\,.
    \end{split}
\end{align}

\section{Consistency check of the anti-hard-collinear jet functions at leading order}
\label{consistencycheck}
We have introduced two definitions of the anti-hard-collinear jet function: one arising within the factorisation formula, see~\eqref{antijet1}, and one obtained via matching, see~\eqref{antijet2}. The first definition is based on the Fierz-transformed operator $Q^q_{1,2}$, the second on the original one. It is well known that the renormalisation properties of the two operators differ starting at $\mathcal{O}(\alpha_s)$, but this issue is well under control (see~\cite{Aebischer:2023djt} and references therein). At leading order, however, the two definitions must lead to the same result.

At leading order, the anti-hard-collinear jet function computed from~\eqref{antijet1} reads
\begin{align}
J_{\text{LO}}^{(\bar n)} =\frac{E_\gamma}{4\pi^2} \frac{1}{\omega_1 + i0}
\left[
1
-\eps\left(
\ln\!\left(\frac{-2E_\gamma\omega_1}{\mu^2}\right)
-2
\right)
\right]
+\mathcal{O}(\eps^2)\,.
\end{align}
This function can also be split into two parts, $J^{(\bar n)}_{\text{LO}}=J^1_{\text{LO}}+J^2_{\text{LO}}$, corresponding to the two derivative terms in the last line of~\eqref{antijet1}:
\begin{align}
J^1_{\text{LO}}=\frac{E_\gamma}{4\pi^2}\frac{1}{\omega_1+i0}
\left[
1
-\eps\left(
\ln\!\left(\frac{-2E_\gamma\omega_1}{\mu^2}\right)
-3
\right)
\right]
+\mathcal{O}(\eps^2)\,,\,\,\,\,\, J^2_{\text{LO}}=-\frac{E_\gamma}{4\pi^2}\frac{\eps}{\omega_1+i0}
+\mathcal{O}(\eps^2)\,.
\end{align}
As in the matching approach, one of the two contributions is evanescent.

To obtain the anti-hard-collinear jet function from the matching condition~\eqref{antijetmatch}, we need the $b \to s \gamma g$ amplitude induced by the operator $Q_1^q$. Expanding the LO result without $u$-dependence, given in~\eqref{eq:one_loop_massless_result_finite}, in $\eps$, we obtain
\begin{align}
\mathcal{A}_{\text{LO}}(b\to s\gamma g)
=&\,
\frac{g_sQ_u E_\gamma}{8\pi^2}
\Bigg[
1
-\eps\left(
\ln\!\left(\frac{-E_\gamma\omega_1}{\mu^2}\right)
+\ln2
-4
\right)
\Bigg]
\nonumber\\
&\times
\Bar{u}^{(s)}_{hc,L}
\left[
\slashed{\epsilon}_{\perp}^*(k_\gamma),
\slashed{\epsilon}_{s,\perp}^*(k_g)
\right]
\frac{\slashed{\bar n}}{2}
b_L
+\mathcal{O}(\eps^2)\,.
\end{align}
Using the definition~\eqref{antijet2}, $J^{(\bar{n})} =
\kappa\left[(2-2\eps)J^{(\bar{n}),a}+2\eps J^{(\bar{n}),b}\right]$, including the factor $\frac{1}{\omega_1+i0}$ discussed in Section~\ref{sec:antijet_from_matching}, we find
\begin{align}
J^{(\bar n)}_{\text{LO}}  =\,\, \kappa\frac{ E_\gamma}{4\pi^2}\frac{1}{\omega_1+i0}
\left[
1
-\eps\left(
\ln\!\left(\frac{-2E_\gamma\omega_1}{\mu^2}\right)
-2
\right)
\right]
+\mathcal{O}(\eps^2)\,,
\end{align}
which agrees with the leading-order result obtained from~\eqref{antijet1} for $\kappa=1$. Note that the couplings $g_sQ_u$ have been factored out in the matching condition~\eqref{antijetmatch}. The additional factor $1/(\omega_1+i0)$ arises because, as explained in Section~\ref{sec:antijet_from_matching}, the soft gluon field is expressed in terms of the field-strength tensor $G_s^{\mu_\perp\nu}(x_+)$ in the factorisation formula, whereas the matching condition uses the standard SCET field $\mathcal{A}^\mu_\perp(x_+)$.
At leading order, the individual contributions are related by
\begin{equation}
J^1_{\text{LO}} = \frac{2 -2\eps}{\omega_1 +i0}\, \, J^{(\bar n), a}_{\text{LO}} +\mathcal{O}(\eps^2)\,,\,\,\,\,\, J^2_{\text{LO}} = \frac{2\eps}{\omega_1 +i0}\, \, J^{(\bar n), b}_{\text{LO}}+\mathcal{O}(\eps^2)\,.
\end{equation}
Finally, all results in this appendix are consistent with the leading-order analysis of~\cite{Benzke:2010js}.
\end{appendix}

\newpage 

\bibliographystyle{JHEP}
\bibliography{bibliography}

 \end{document}